\documentclass[12pt]{article}

\usepackage{aaspp4}
\usepackage{amsfonts}

\usepackage{aabib}

\usepackage{graphicx}

\newcommand{\mH}{\ensuremath{m_{\mbox{\scriptsize\rm p}}}}
\newcommand{\NT}{\ensuremath{N_{\mbox{\scriptsize\rm T}}}}
\newcommand{\Lmech}{\ensuremath{L_{\mbox{\scriptsize\rm mech}}}}
\newcommand{\Rsb}{\ensuremath{R_{\mbox{\scriptsize\rm sh}}}}
\newcommand{\del}{\ensuremath{{d}}}
\newcommand{\Msun}{\ensuremath{\rm M}_\odot}
\newcommand{\Msolar}{\ensuremath{\rm M}_\odot}
\newcommand{\mach}{\ensuremath{\cal M}}
\newcommand{\erf}{\ensuremath{{\rm erf}}}
\newcommand{\rcl}{\ensuremath{r_{\mbox{\scriptsize\rm cl}}}}
\newcommand{\Nc}{\ensuremath{N_{\mbox{\scriptsize\rm cl}}}}
\newcommand{\rc}{\ensuremath{r_{\mbox{\scriptsize\rm cl}}}}
\newcommand{\Vc}{\ensuremath{V_{\mbox{\scriptsize\rm cl}}}}
\newcommand{\ncl}{\ensuremath{n_{\mbox{\scriptsize\rm cl}}}}
\newcommand{\nc}{\ensuremath{n_{\mbox{\scriptsize\rm cl}}}}
\newcommand{\acl}{\ensuremath{a_{\mbox{\scriptsize\rm cl}}}}

\newcommand{\deltap}{\ensuremath{\delta_{\mbox{\scriptsize\rm p}}}}
\newcommand{\etaesc}{\ensuremath{\eta_{\mbox{\scriptsize\rm esc}}}}
\newcommand{\thetac}{\ensuremath{\theta_{\mbox{\scriptsize\rm c}}}}
\newcommand{\nsh}{\ensuremath{n_{\mbox{\scriptsize\rm sh}}}}
\newcommand{\Rtr}{\ensuremath{R_{\mbox{\scriptsize\rm tr}}}}
\newcommand{\nH}{\ensuremath{n_{\mbox{\scriptsize\rm H}}}}

\newcommand{\Zacc}{\ensuremath{Z_{\mbox{\scriptsize\rm acc}}}}
\newcommand{\Rcy}{\ensuremath{R_{\mbox{\scriptsize\rm cy}}}}
\newcommand{\Ztrans}{\ensuremath{Z_{\mbox{\scriptsize\rm tr}}}}
\newcommand{\Ztr}{\ensuremath{\Ztrans}}
\newcommand{\Zcy}{\ensuremath{Z_{\mbox{\scriptsize\rm cy}}}}
\newcommand{\Sesc}{\ensuremath{S_{\mbox{\scriptsize\rm esc}}}}

\newcommand{\thetadyn}{\ensuremath{\theta_{\mbox{\scriptsize\rm dyn}}}}
\newcommand{\sigmah}{\ensuremath{\sigma_{\mbox{\scriptsize\rm h}}}}
\newcommand{\sigmat}{\ensuremath{\sigma_{\mbox{\scriptsize\rm t}}}}
\newcommand{\Nass}{\ensuremath{N_{\mbox{\scriptsize\rm as}}}}
\newcommand{\dotNass}{\ensuremath{\dot{N}_{\mbox{\scriptsize\rm as}}}}

\newcommand{\Snot}{\ensuremath{S_{\mbox{\scriptsize\rm 0}}}}

\newcommand{\tmin}{\ensuremath{t_{\mbox{\scriptsize\rm 1}}}}
\newcommand{\tmax}{\ensuremath{t_{\mbox{\scriptsize\rm 2}}}}
\newcommand{\Psilyc}{\ensuremath{\Psi_{\mbox{\scriptsize\rm LyC}}}}
\newcommand{\Psiesc}{\ensuremath{\Psi_{\mbox{\scriptsize\rm esc}}}}

\newcommand{\dotNstar}{\ensuremath{\dot{N}_{\star}}}

\newcommand{\gta}{\gtrsim}
\newcommand{\lta}{\lesssim}
\newcommand{\avg}[1]{\ensuremath{\langle#1\rangle}}

\newcommand{\fesc}{\ensuremath{\avg{f_{\mbox{\scriptsize\rm esc}}}}}

\newcommand{\ltsima}{$\; \buildrel < \over \sim \;$}
\newcommand{\simlt}{\lower.5ex\hbox{\ltsima}}
\newcommand{\gtsima}{$\; \buildrel > \over \sim \;$}
\newcommand{\simgt}{\lower.5ex\hbox{\gtsima}}

\begin{document}

\title{The Escape of Ionizing Photons from OB Associations in Disk Galaxies:
Radiation Transfer Through Superbubbles}

\author{James B. Dove$^{1,2}$, J. Michael Shull$^{1,3}$, and Andrea
Ferrara$^{4,5}$}
\centerline{$^1$ CASA, Department of Astrophysical and Planetary Sciences,}
\centerline{University of Colorado, Campus Box 389, Boulder, CO 80309}
\centerline{$^2$ Also at Department of Physics, }
\centerline{Metropolitan State College of Denver, Campus Box 69, Denver, CO 80217}
\centerline{$^3$ Also at JILA,Campus Box 440, Boulder, CO 80309}
\centerline{$^4$ Osservatorio Astrofisico di Arcetri, I-50125 Firenze, Italy}
\centerline{$^5$ JILA Visiting Fellow}

\begin{abstract}
By solving the time-dependent radiation transfer problem of stellar
radiation through evolving superbubbles within a smoothly varying H~I
distribution, we estimate the fraction of ionizing photons emitted by OB
associations that escapes the H~I disk of our Galaxy into the halo and
intergalactic medium (IGM).  We consider both coeval star-formation and a
Gaussian star-formation history with a time spread $\sigmat = 2$ Myr.  We
consider both a uniform H~I distribution and a two-phase (cloud/intercloud)
model, with a negligible filling factor of hot gas.  We find that the
shells of the expanding superbubbles quickly trap or attenuate the ionizing
flux, so that most of the escaping radiation escapes shortly after the
formation of the superbubble. Superbubbles of large associations can blow
out of the H~I disk and form dynamic chimneys, which allow the ionizing
radiation to escape the H~I disk directly. However, blowout occurs when the
ionizing photon luminosity has dropped well below the association's maximum
luminosity.  For the coeval star-formation history, the total fraction of
Lyman Continuum photons that escape both sides of the disk in the solar
vicinity is $\fesc \approx 0.15 \pm 0.05$. For the Gaussian star formation
history, $\fesc \approx 0.06 \pm 0.03$, a value roughly a factor of two
lower than the results of \cite*{dove94b}, where superbubbles were not
considered.  For a local production rate of ionizing photons $\Psi_{\rm
LyC} = 4.95\times 10^7$ cm$^{-2}$ s$^{-1}$, the flux escaping the disk is
$\Phi_{\rm LyC} \approx (1.5-3.0)\times 10^6$ cm$^{-2}$ s$^{-1}$ for coeval
and Gaussian star formation, comparable to the flux required to sustain the
Reynolds layer.  Rayleigh-Taylor instabilities exist early in the OB
association's evolutionary stages, possibly causing the shell to fragment
and increasing $\fesc$. However, if a significant fraction of H~I is
distributed in cold clouds with $\nH \sim 30$ cm$^{-3}$, $\fesc$ can be
reduced by a factor of $\sim 2-5$ if the cloud properties are similar to
``Standard Clouds'' with a disk geometry.
\end{abstract}


\setcounter{footnote}{0}

\section{Introduction}\label{sec:intro}
The diffuse, ionized medium (DIM), a.k.a the ``Reynolds layer,'' has been
established to occupy a significant fraction of the interstellar medium
(ISM) and to have a vertical scale height of $\sim$ 1 kpc
(\cite{mezger78a,reynolds91a,reynolds91b}).  It is widely believed that OB
associations are the sources of radiation responsible for ionizing the
Reynolds layer (\cite{reynolds84a,mathis86a,bregman86a}).  It is still
unclear how this ionizing radiation is able to travel so far from the OB
associations' immediate vicinity.  In a previous paper (\cite{dove94b},
hereafter referred to as DS94), we calculated the geometry of diffuse H~II
regions due to OB associations in the Galactic plane by assuming a
three-component model (\cite{dickey90a}) for the vertical distribution of
hydrogen. For associations producing an ionizing luminosity $\Snot \gta
3\times 10^{49}$ s$^{-1}$, we found that the H~II regions are ``density
bounded'' in the vertical direction, allowing photons to escape into the
halo of our Galaxy. By integrating over the luminosity function of
associations, $\del N_a(\Snot)/\del\Snot$ (\cite{kennicutt89a,mckee97a}),
we estimated that roughly 7\% of ionizing photons, corresponding to a
number flux $\Phi_{\rm LyC} = 2 \times 10^6$ cm$^{-2}$ s$^{-1}$, escape
each side of the H~I disk layer in the solar vicinity and penetrate the DIM
layer. If this radiation is absorbed in the DIM, assumed to lie entirely
above the HI layer, the corresponding
photoionization rate would be comparable to the recombination rate implied
by the H$\alpha$ emission measurements
(\cite{mezger78a,reynolds91a,reynolds91b}).  If roughly one-third of this
radiation escapes the Galactic halo, this flux is also consistent with the
estimated flux required to photoionize the Magellanic Stream and several
high velocity clouds in the solar vicinity (\cite{bland-hawthorn99a}). 
Rather than assuming a purely diffuse ISM, \cite*{miller93a} considered the
effects of a statistical distribution of small opaque clouds, also
decreasing with increasing Galactic height and embedded within the diffuse
gas. These authors found that the resulting H~II regions are consistent
with the observations of dispersion measures and H$\alpha$ emission
measurements.

Although these results are encouraging, no models to date consider the role
of dynamic chimneys and superbubbles produced by the OB associations. These
considerations are the focus of this paper.  We recently became aware of a
preprint on a similar topic (\cite{basu99a}).  For an OB association, the
stars that produce most of the ionizing radiation have strong stellar winds
and form supernovae after their relatively short lifetime. Therefore,
shortly after the formation of an OB association, a superbubble will form,
with a thin shell behind the shock front excavating a large cavity of hot
gas (\cite{weaver77a,mccray79a,mccray87a,shull95a}). For a sufficiently large 
association, the superbubble can break out of the H~I disk, forming a
dynamic chimney, where the hot cavity penetrates into the halo
(\cite{maclow88a}).

As we discuss below, the shells of the expanding superbubbles quickly trap
the ionizing radiation, so that no radiation can escape the disk until
blowout.  After blowout, a large fraction ($\sim$ 10\% per side) of
ionizing radiation can directly escape the disk of the galaxy by
propagating through the dynamic chimney. However, this is a viable
mechanism for transporting the ionizing radiation only if an appreciable
amount of ionizing radiation is still being produced within the cavity
after the dynamic chimney has formed.  We find that, for both a coeval star
formation history or a non-coeval model having a star formation rate
given by a Gaussian with a full-width half-maximum $\sigma_t = 2$ Myr
(\cite{massey98a,garmany98a}), a small fraction of the ionizing radiation
produced by an association over its lifetime is emitted after the time of
blowout. Therefore, most of the radiation that escapes the disk does so
early in the early stages of the superbubble. Integrating over the
luminosity function, we find that the fraction of ionizing photons {\em
currently} being emitted by OB associations that escape each side of the
H~I disk is $\fesc \sim 15$\% for the coeval star-formation history and
$\fesc \sim 6$\% for the Gaussian star-formation history.

The aim of this paper is to predict the fraction of ionizing radiation,
emitted by OB associations, that escapes the H~I disk of the Galaxy. As
discussed above, DS94 assumed that the Reynolds layer lies above the H~I
layer. Therefore, the question of whether the Reynolds layer is
sustained via hot star photoionization is equivalent to whether
enough radiation can penetrate through the H~I layer. In this paper, we
regard this assumption as too simplistic. The Reynolds layer
has a scale height of roughly 1 kpc, whereas the H~I disk has a scale
height of $\sim 200$ pc (\cite{dickey90a}) but extends to at least 1 kpc
(\cite{lockman86}).  Thus, a significant fraction (e.g., $\sim 20$\%) of 
the of the ionized gas may lie within the H~I layer,
reducing the flux of ionizing radiation required to escape the H~I disk. A
self-consistent model, modeling the distribution of the photoionized gas,
is not feasible until the density distribution of the hydrogen gas, both
within the disk and in the halo, is determined more accurately.

In \S2, we discuss our treatment of the evolution of the superbubble and
the radiation transfer of ionizing photons emitted by an association
through the expanding superbubble and the diffuse ISM of the disk. We also
discuss our treatment of the blowout event of the superbubble, where the
cavity evolves into a dynamic chimney. In \S3, we determine the time
dependence of the fraction of ionizing photons, emitted by a single OB
association, that escape the H~I disk.  Integrating over the luminosity
function of OB associations, we determine the fraction of ionizing
radiation, currently being emitted by OB associations, that escapes the H~I
disk. In \S4, we discuss key issues not included in our standard model,
including shell instabilities, a two-phase ISM, ``poisoning'' by
photoablated gas, and triggered star formation. We then discuss how these
effects may alter our results. In \S5, we discuss the implications of our
results and give our conclusions.

\section{Radiation Transfer Through an Expanding Superbubble}
\subsection{Evolution of the Superbubble}\label{sec:superbubble}
Consider a large association having $\NT$ stars in the $8-85\ \Msun$ mass
range. Due to the stellar winds from the massive stars and type II
supernovae, the association produces a mechanical luminosity, \Lmech$(t)$,
that varies with time. This energy input drives an expanding
superbubble. Assuming an initially uniform density distribution of the
diffuse ISM, and that all of the swept-up mass resides in a thin
shell ( the Kompaneets approximation, \cite{kompaneets60a}), the radius,
\Rsb, of the cavity is given by (\cite{shull95a})
\begin{equation}\label{eq:kompaneets}
\frac{\del}{\del t}\left[\Rsb^3\frac{\del^2}{\del
t^2}\left(\Rsb^4\right)\right] = \frac{6 \Rsb^2}{\pi\rho}\Lmech(t),
\end{equation}
where the mass density $\rho = 1.4 \mH n_H$, $n_H$ is the number density of
hydrogen, and $\mH$ is the proton mass. For a given time-dependent
mechanical luminosity prescription, this equation is integrated numerically
using a fourth-order Runge-Kutta algorithm. We note that this formalism is
technically valid only for a uniform density distribution, as the shock
wave does not evolve with spherical symmetry for non-uniform density
distributions.  However, as we show in \S3, for most association sizes the
expanding shell becomes optically thick to the ionizing radiation before
the shock has propagated a distance of a disk scale height, within which
using a constant density is a good approximation. Therefore, the amount of
escaping ionizing radiation is insensitive to the errors produced by
neglecting the non-spherical propagation that occurs at large Galactic
heights.

The time dependence of the mechanical luminosity depends on the initial
mass function (IMF) of stars within an OB association and the star
formation history of the association. For this paper, we use the OB
association evolution models of \cite*{sutherland99a}, which determine both
the mechanical luminosity and the luminosity of ionizing photons as a
function of time for individual stars from the latest stellar evolutionary
tracks and stellar atmospheres. These authors used an IMF given by $\del
N/\del M \propto M^{-(\gamma + 1)}$, where $\gamma = 1.6$ for the stellar
mass range $8\ \Msolar \le M \le 85\ \Msolar$ (\cite{garmany98a}), for which
all stars eventually become supernovae.
The massive star-formation history of the
association is given by $\dotNstar(t)$, with the constraint that $\int
\dotNstar(t) \del t = \NT$, where the limits of integration span the entire
lifetime of the association. In this paper, we consider two star-formation
histories (SFHs): (1) coeval [$\dotNstar(t) \propto \delta(t-t_o)$], and
(2) noncoeval with a Gaussian distribution [$\dotNstar(t) \propto
\exp(-t^2/2 \sigma_t^2)]$.  For the Gaussian SFH, the IMF is assumed to be
constant in time.

For each SFH, $\Lmech(t)$ and $S(t)$ were obtained from
\cite*{sutherland99a}. Here, we used a Monte-Carlo simulation to
determine the mass distribution of individual associations having a total
mass of $10^4\ \Msolar$ (corresponding to an average association size of
$3000$ stars, or roughly $\NT\sim 300$ massive stars). The average
luminosity curves are then normalized, and, for an arbitrary association
size, the luminosity curves are given by the product of the normalized
curves and \NT. Note that this formalism ignores all possible deviations
about the average luminosities due to statistical fluctuations of the
stellar mass distribution of small associations. \cite*{sutherland99a}
found that these fluctuations become significant for $\NT \lta
100$. However, as shown below, most of the escaping radiation in our Galaxy
emanates from associations with $\NT \gta 100$, so the importance of the
statistical fluctuations of small associations is minor for our
purposes. Even if the small associations were important, when calculating
the total fraction of escaping photons, we integrate over many
associations, so the use of the average luminosities should yield a good
approximation.

\begin{figure}
\centerline{\includegraphics[width=.5\textwidth]{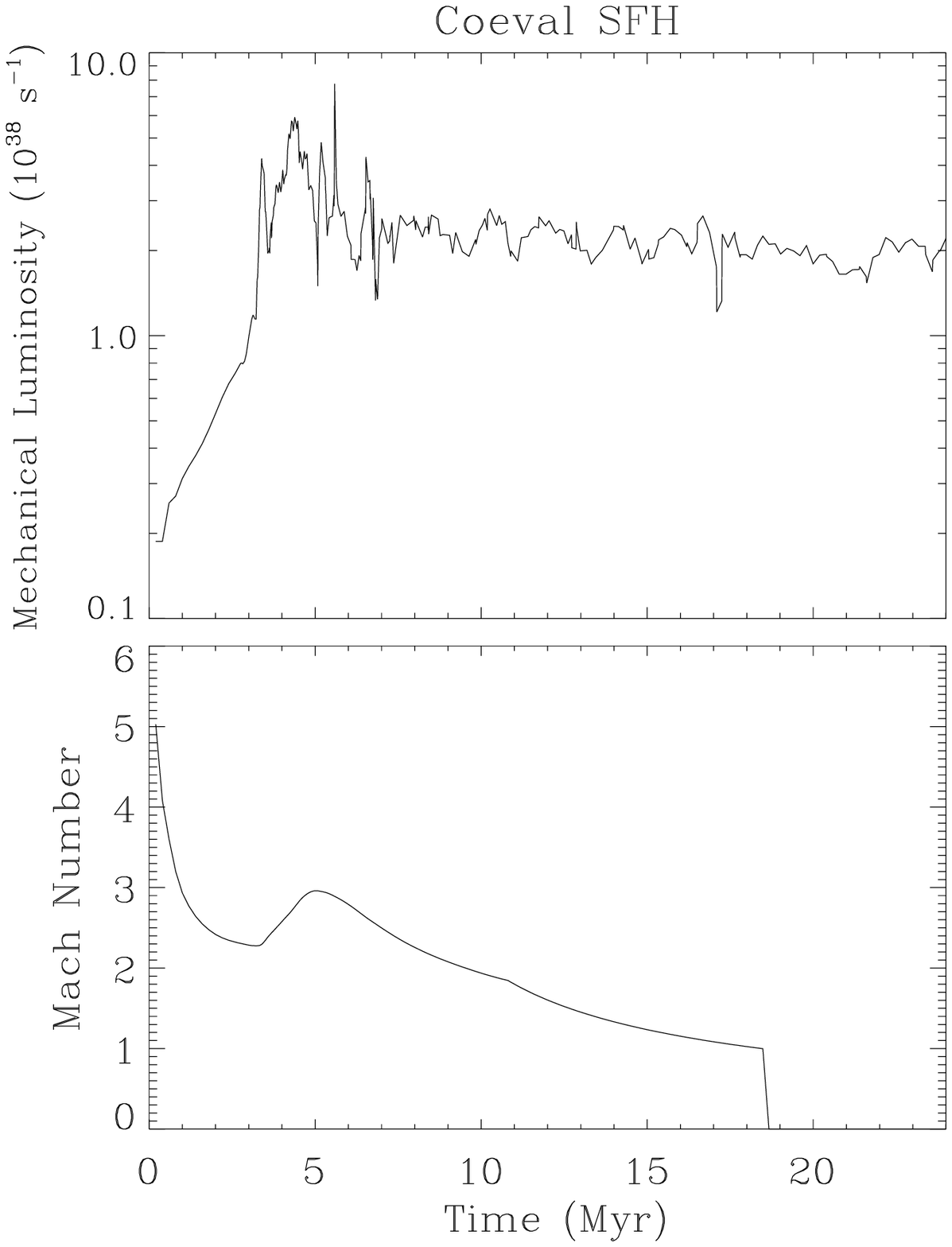}
\hfill%
\includegraphics[width=.5\textwidth]{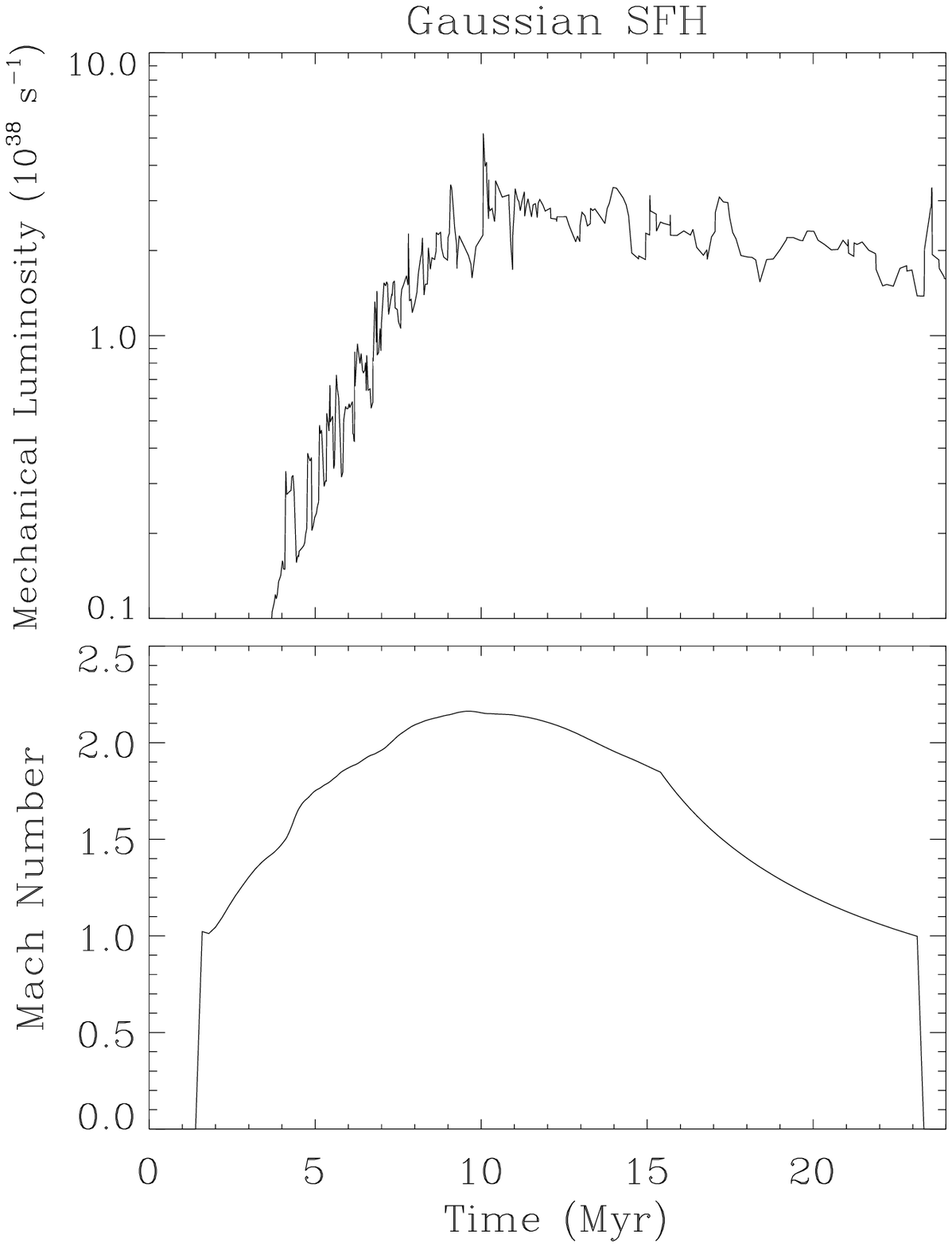}}
\figcaption{Mechanical luminosity of the OB association and Mach number of the
shell as a function of time. For the Gaussian SFH, $\sigma_t=2$ Myr and the
star-formation peak occurs at $t=2\sigma_t$. For both cases, $\NT = 200$
and $\beta = 1$.  Blowout occurs when $t\approx 11$ Myr for the coeval SFH
and $t\approx 16$ Myr for the Gaussian SFH. After blowout, the shell is in
the momentum-conserving phase, and the Mach number shown is evaluated at
the midplane.
\label{fig:Lvst}}
\end{figure}

In Figure \ref{fig:Lvst}, we show the mechanical luminosity and the Mach
number of the shell as a function of time for an OB association having $\NT
= 200$, for coeval star-formation history and for a Gaussian star-formation
history with $\sigma_t = 2$ Myr.  The Mach number of the shock front is
${\cal M} = v_{\rm sh}/c_{\rm s}$, where the effective sound speed of the
medium is
\begin{equation}
c_{\rm s} = \sqrt{\frac{kT}{\mu_{\rm p}\mH}(1 + \beta)}.
\end{equation}
Here, $k$ is the Boltzmann constant, $T$ is the temperature of the gas
(assumed equal to $10^4$ K and the shock front is assumed to be
isothermal), $\mu_{\rm p} = 0.6$, and $\beta$ is the ratio of the magnetic
pressure plus turbulent pressure to the thermal pressure.  As discussed in
\S\ref{sec:blowout}, the superbubble is modeled as instantaneously blowing
out of the disk and becoming a dynamic chimney when the bubble radius
exceeds the transition height, $\Ztrans$. We assume that $\Ztrans$ is
twice the scale height of the H~I gas distribution, $\sigmah$. After
blowout, the cylindrical radius continues to expand due to conservation of
momentum. Once the Mach number reaches unity, the expansion of the cavity
is halted.

The fluctuations in the mechanical luminosity arise from supernovae
occurring separately in time due to the different lifetimes of individual
stars. Although $\Lmech(t)$ is given by the average of several Monte-Carlo
simulations, the fluctuations have not been smoothed out.

We note that, for the Gaussian SFH, the Mach number is near zero until 1.3 Myr,
when it then rises quickly. This behavior is due to our initial treatment
of the bubble growth for low mechanical luminosities. Since we start the
simulation at a time $2\sigma_t$ before the peak star-formation rate, the
initial luminosity is small for small associations. In such cases,
the Mach number rapidly drops below unity, then subsequently increases
as the energy input increases. In reality, the shock front associated
with these small associations would stagnate when the ambient pressure of
the ISM is equal to the internal pressure of the bubble, causing the
Kompaneets approximation to break down. As the proper hydrodynamic
treatment of the growth of a stagnated bubble is outside the scope of this
paper, we employed the following procedure for approximating the bubble
evolution. For each association size, we determine the time at which the
mechanical luminosity exceeded $10^{36}$ erg s$^{-1}$. We then evaluate the
Mach number at this time using the solution of equation
(\ref{eq:kompaneets}) for a constant luminosity, where the time-averaged
luminosity is used. If the Mach number is below unity, we assume that the
shock has stagnated, set the shell velocity to zero, and continue
integrating equation (\ref{eq:kompaneets}). Although the use of Kompaneet's
equation is not valid for Mach numbers below unity, we argue that the
errors introduced are negligible since the time interval for the shell to
accelerate to Mach numbers above unity is small relative to the
sound-crossing time of the bubble. Therefore, the shell does not diffuse
much into the cavity during the stagnation period.  Furthermore, during
this early interval of the cavity, the photon luminosity of the association
is also low, and no radiation escapes the disk of the galaxy regardless of
the bubble geometry. Finally, by comparing ${\cal M}(t)$ using this method
with results when equation (\ref{eq:kompaneets}) is integrated without
intervention, we found that the two were identical for $t \gta 1.5$ Myr. The
only ramification of our stagnation treatment is that smaller
associations take longer to blow out of the disk. As discussed in
\S\ref{sec:instabilities}, the acceleration of stagnated bubbles due to the
mechanical luminosity increasing in time can cause Rayleigh-Taylor
instabilities to begin at a much earlier stage, compared to the case of
constant luminosity, where acceleration occurs at the time of blowout
(\cite{maclow88a}).  This instability may dramatically increase the amount
of ionizing radiation that escapes the H~I disk.

\subsection{Superbubble Blowout}\label{sec:blowout}
For large OB associations that have a sufficiently high mechanical
luminosity, superbubbles can blow out of a vertically stratified H~I disk,
producing dynamic chimneys
(\cite{heiles79a,tomisaka86a,maclow88a,maclow89a}).  Blowout occurs if the
shell of the superbubble is still supersonic when it reaches a height,
$\Zacc$, where it begins to accelerate due to decreasing density
distribution (\cite{maclow88a}). This height depends on the density
distribution of the ISM and $\Lmech$; for Gaussian or exponential
distributions it is $\Zacc \sim (2-3)\sigmah$ (\cite{maclow88a,maclow99a}).
For most association sizes and star formation histories, the timescale for
reaching \Zacc\ is considerably longer than the timescale for propagating
from this height into the Galactic halo ($Z \gta 1$ kpc).  Therefore,
rather than calculating the exact evolution of the superbubble during
blowout, we model the process as an instantaneous transition from a bubble
into a dynamic chimney once the shell reaches a transition height $\Ztrans
=\Zacc$.  We note that, due to the existence of the H~I gas at high $Z$,
blowout may first occur at heights considerably larger than $3\sigmah$ or
may not occur at all. In addition, magnetic fields could inhibit blowout
(\cite{tomisaka98a}).  As discussed below, our results are insensitive to
the exact value of \Ztrans, since most ionizing radiation that escapes the
disk does so shortly after the formation of the OB association. Therefore,
the uncertainties regarding the process of blowout are unimportant for our
estimation of the flux of escaping ionizing radiation.

The geometry of the dynamic chimney is approximated as a 
cylinder, centered on the OB association, with radius \Rcy\ and height
\Zcy.  Since we are only interested in how much ionizing radiation escapes
the H~I disk of the Galaxy, we do not attempt to model the structure of the
superbubble/chimney for heights more than $\sim 4-5$ disk scaleheights
($\gta 1$ kpc). Numerical simulations of superbubbles blowing out of
stratified disks (\cite{tomisaka98a,maclow89a,mineshige93a}) show that a
cylinder is a decent approximation for the geometry of the cavity within
the H~I disk. Above the disk, the shell takes the appearance of a mushroom
cloud, with the shell clumping and fragmenting due to Rayleigh-Taylor
instabilities. For this paper, we assume that any radiation that reaches a
height \Zcy\ and is within the dynamic chimney escapes the disk. The
question of whether this radiation is absorbed by the ``mushroom cap'' of
the shell or escapes due to the shell having a small covering fraction is
outside the scope of this paper. Even if the covering fraction of the
``mushroom cap'' is not negligible, it is possible that this gas,
photoionized by the OB association, is part of the Reynolds layer.
Therefore, it is fair to include this radiation when assessing whether
the amount of escaping radiation can ionize the DIM.

%
%
\subsection{Ionization Equilibrium of the H~II Region}

Throughout this paper, we assume that ionization equilibrium is satisfied
instantaneously.  As discussed in DS94,
 the ionization front in a Gaussian disk layer will reach a
Galactic height $Z$ in a time
\begin{eqnarray}
t(Z) &=&
15.5\left[\frac{\sqrt{\pi}}{4}\erf\left(\frac{Z}{\sqrt{2}\sigmah}\right)\right. \\  
&&- \left. \frac{Z}{2^{3/2}\sigmah}\exp\left(-\frac{Z^2}{2\sigmah^2}\right)\right]\nonumber\\
&& \times \left(\frac{\sigmah}{0.25\ {\rm
kpc}}\right)^3\left(\frac{n_0}{0.3\ {\rm cm}^{-3}}\right) S_{\rm 49}^{-1}\
{\rm Myr} \nonumber,
\end{eqnarray}
where $S_{\rm 49}$ is the photon luminosity is units of $10^{49}$ s$^{-1}$,
$\sigmah$ is the scaleheight of Gaussian disk, and $n_0$ is the number
density of hydrogen in the midplane of the disk. (This equation should
replace equation (3) of DS94, whose equation (3) contains an error.)  For
$S_{\rm 49} = 1$, $\sigmah = 0.184$ kpc, and $n_0 = 0.367$ cm$^{-3}$, the
I-front reaches a height $Z=\sigmah$ in a time $t\approx 1.4$ Myr and
reaches a height $Z=2 \sigmah$ in a time $t\approx 3.2$ Myr, both
relatively small compared to the lifetime of the superbubble. As shown in
DS94, ionization equilibrium is reached within a decade or so after the
passing of the I-front. [For gas with an ionization fraction $X \sim 1$,
ionization equilibrium is reached in a time $\tau \sim (1-X)/(n_{\rm e}
\alpha_{\rm H}^{(2)})$].  Thus, even though the density distribution of the
superbubble is evolving in time (as discussed below), the H~II region is
determined in a quasi-static fashion.

\subsubsection{Geometry of the H~II Regions}
We assume that the stars of individual OB associations are compactly
distributed so that the radiation can be treated as emanating from a point
source.  We define $y(t)$ to be the dimming function of the OB association,
which relates the time-dependent photon luminosity of Lyman-continuum (LyC)
photons and the maximum luminosity via
\begin{equation}
S(\NT,t) = \Snot(\NT) y(t).
\end{equation}
For a point source of ionizing radiation, the H~II/H~I boundary 
is given implicitly by
\begin{equation}
\frac{S(\NT,t)}{4\pi\alpha^{(2)}_H} = \int\limits_{0}^{\Rsb}R^2
n_H^2(R,\theta) \del R,
\end{equation}
where $\theta$ is the angle between the normal of the Galactic disk and the
radial unit vector, $\alpha^{(2)}_H$ is the case-B recombination rate for
hydrogen, and $n_H$ is the number density of hydrogen. Azimuthal symmetry
has been assumed.

For the vertical distribution of H~I, DS94 
considered both the three-component Dickey-Lockman model
(\cite{dickey90a}) as well
as several single Gaussian distributions (with different amounts
of dark matter) which were fitted to numerical calculations assuming
hydrostatic equilibrium (\cite{dove94a}). We found that all models
predicted the same fraction of escaping photons within 10\%.
This Gaussian distribution which most closely resembled the
Dickey-Lockman model is given by
\begin{equation}
n_H(Z) = n_0 \exp(-Z^2/2\sigma_h^2),
\end{equation}
where $Z$ is the vertical distance above the mid-plane of the disk, $n_0 =
0.367$ cm$^{-3}$ and $\sigma_h = 0.184$ kpc.  In this paper, we only
consider this single-Gaussian distribution.
 
As the superbubble expands into the diffuse ISM, all of the swept-up mass
is assumed to reside in the thin shell.  The thickness of the shell, $\Delta
R$, is given by equating column densities,
\begin{equation}
\nsh(\Rsb,\mu) \Delta R = \int\limits_{0}^{\Rsb} n_H(Z) \del R,
\end{equation}
which yields
\begin{equation}
\Delta R = \frac{n_0}{\nsh} \sqrt{\frac{\pi}{2}}
 \left(\frac{\sigma_h}{\mu}\right) \erf\left(\frac{\Rsb
 \mu}{\sqrt{2}\sigma_h}\right),
\end{equation}
where $\nsh$ is the number density within the shell, assumed to be a
constant with respect to radius, and $\mu = \cos\theta$.  Defining the
transition radius $\Rtr = \Rsb + \Delta R$, the geometry of the H~II region
for an OB association with luminosity $S$, situated on the mid-plane of the
Galaxy, is given implicitly by
\begin{eqnarray}\label{eq:HIIgeom}
\frac{S}{4\pi\alpha^{(2)}_H} &=& \frac{1}{3}\nsh^2(\Rsb,\mu)(\Rtr^3 -
\Rsb^3) + \\ &&
n_0^2\left(\frac{\sqrt{\pi}}{4}\left(\frac{\sigma_h}{\mu}\right)^3\left[\erf\left(\frac{\mu\Rsb}{\sigma_h}\right)
- \erf\left(\frac{\mu \Rtr}{\sigma_h}\right)\right]\right. \nonumber \\
 &&+ \left.\frac{\sigma_h^2}{2\mu^2}\left\{\Rtr\exp\left[-\left(\frac{\mu
\Rtr}{\sigma_h}\right)^2\right] - \right.\right.\nonumber \\
&& \left.\left. \Rsb\exp\left[-\left(\frac{\mu\Rsb}{\sigma_h}\right)^2\right]\right\}
\right)\nonumber
\end{eqnarray}
The density of this shell is given by the isothermal shock condition for
compression in the cool post-shock layer,
\begin{equation}
\nsh(\Rsb,\mu) = {\cal M}^2 \nH(\mu\Rsb).
\end{equation}

\begin{figure}
\centerline{\includegraphics[width=0.8\textwidth]{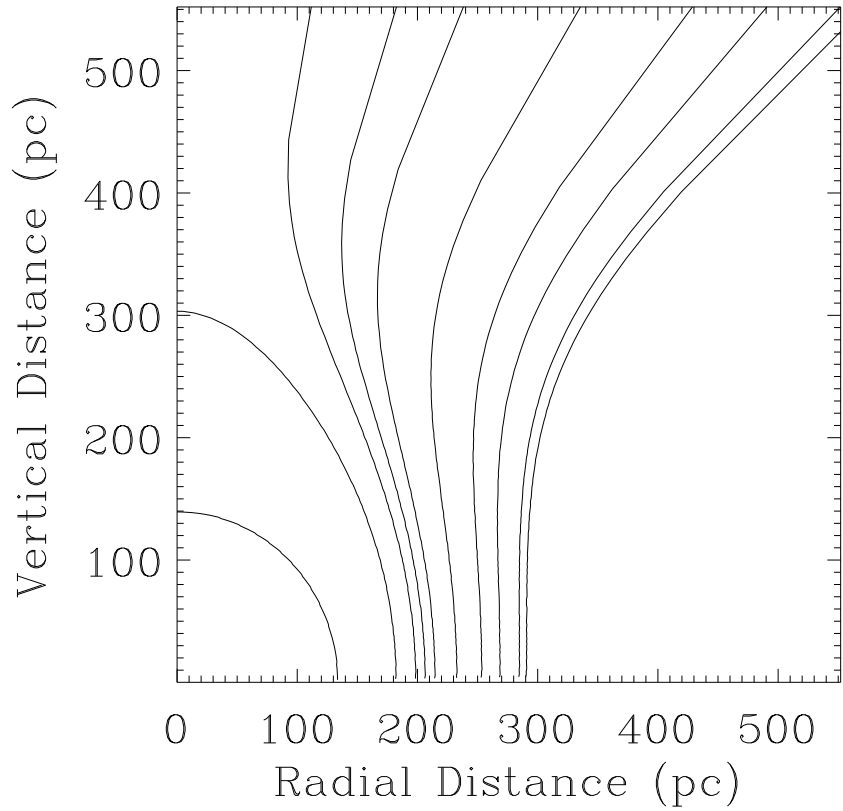}}
\figcaption{Geometry of H~II region for several times after the formation
of a coeval OB association with $\NT = 50$ massive stars. From
right to left, the time (in Myr) after formation is $t = 0.0, 1.6,
2.2, 2.4, 2.6, 2.7, 2.8, 2.85, 2.9,$ and $4.0$.
\label{fig:HIIgeomvst.coeval.N50}}
\end{figure}

\begin{figure}
\centerline{\includegraphics[width=0.8\textwidth]{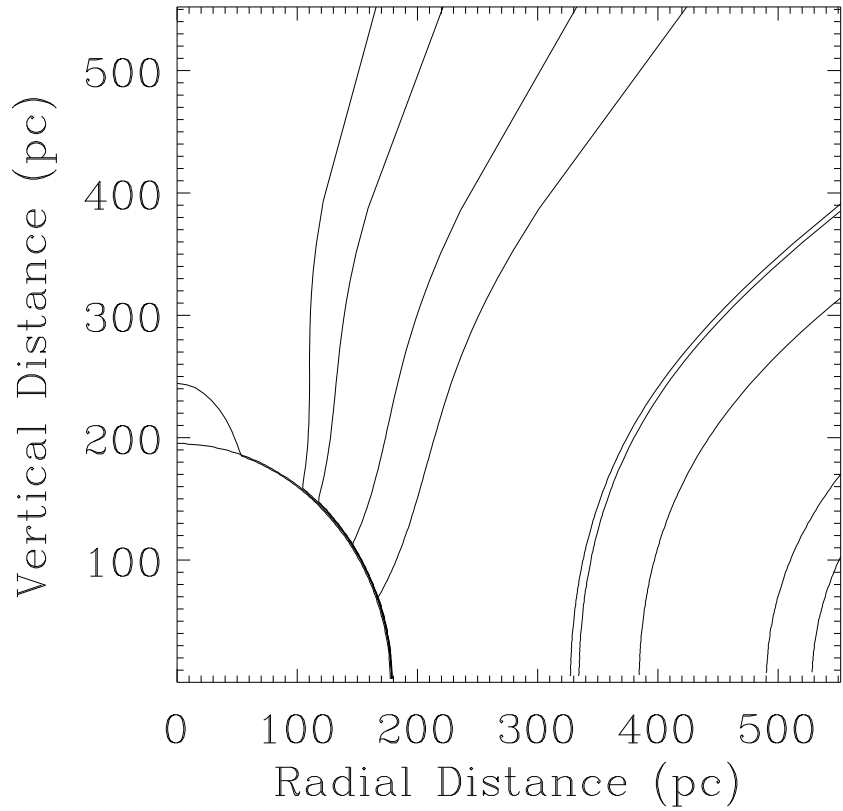}}
\figcaption{Geometry of H~II region for several times after the formation
of a coeval OB association of size $\NT = 300$. From
right to left, the time (in Myr) after formation is $t = 0.0, 2.20,
2.77, 2.93, 3.13, 3.36, 3.40, 3.50, 3.51, 3.54$, and $3.61$.
\label{fig:HIIgeomvst.coeval.N300}}
\end{figure}

In Figures \ref{fig:HIIgeomvst.coeval.N50} and 
\ref{fig:HIIgeomvst.coeval.N300}, we show the time evolution of the geometry
of the H~II region shortly after the formation of an OB association, for
$\NT = 50$ and $\NT = 300$, respectively, assuming coeval star formation.
For both association sizes, the H~II region is initially density bounded in
the vertical direction (see below for discussion of how much radiation
escapes the H~I disk). As the superbubble expands, the column density of
the shell increases. Since the shell has a higher density than the
diffuse gas, and therefore a higher recombination rate, the volume of the
H~II region decreases with time. Eventually, the H~II region is radiation
bounded even in the vertical direction. For $\NT = 300$, a portion of the
H~I/H~II interface lies within the shell for $t\gta 3.0$ Myr, and by $t\sim
3.6$ Myr the entire interface is within the shell. The H~II region goes
from being density limited to radiation limited in a short time
interval.  For $\NT = 50$, the H~II region becomes radiation
limited at only a slightly earlier time compared to the $\NT =
300$ association ($2.9$ Myr compared to $3.5$ Myr) even though its
photon luminosity is six times lower.  This is due to the smaller association
having lower shell velocities, causing both a slower accumulation rate
of the column density as well as lower gas densities within the compressed
shell. 

\subsubsection{Escaping Radiation}\label{sec:etaesc}
If the H~II region is density bounded, then some radiation escapes the H~I
disk. The fraction of ionizing photons that escape along the differential
solid angle $\del\Omega(\theta) = \sin\theta\ \del\theta\del\phi$ is
\begin{eqnarray}
f[\theta,S(\NT,t)] &=& 1 -
\frac{4\pi\alpha^{(2)}_H}{S(\NT,t)}\int\limits_{0}^{\infty}
n_H^2(R,\theta,t) R^2\ \del R\nonumber\\ 
&=& 1 - \frac{4\pi n_0^2\alpha^{(2)}_H}{S(\NT,t)}
\left\{\frac{1}{3}\left(\frac{\nsh}{n_0}\right)^2(\Rtr^3
- \Rsb^3) + \right. \nonumber\\ & & \left.
\left(\frac{\sigma_h}{\mu}\right)^3\left(\frac{\sqrt{\pi}}{4}\left[1-\erf\left(\frac{\mu\Rtr}{\sigma_h}\right)\right]
+\right.\right.\nonumber \\ &&
\left.\left.\frac{\Rtr\mu}{2\sigma_h}\exp\left[-\left(\frac{\mu\Rtr}{\sigma_h}\right)^2
\right]\right)\right\}.
\end{eqnarray}

The critical angle, $\theta_c(S)$, is the angle at which the H~II region is
radiation bounded for $\theta > \theta_c$ and density bounded for $\theta <
\theta_c$. The critical angle is found by solving $f(\theta_c,S) = 0$.  In
contrast to DS94, $\theta_c$ is time-dependent since both the
density distribution and the ionizing luminosity vary with time. Finally,
we define the fraction of photons, emitted by a single OB association
having size \NT, age $t$, and corresponding luminosity $S(\NT,t)$, that
escape {\em each} side of the H~I disk to be
\begin{equation}\label{eq:etaesc}
\etaesc(\NT,t) = \int\limits_{0}^{\thetac}\frac{f(\theta,S)}{4\pi} 2\pi
\sin\theta\ \del\theta.
\end{equation}

\smallskip
\subsubsection{H~II Regions for Dynamic Chimneys}
For a dynamic chimney with height \Zcy\ and cylindrical radius \Rcy, the
angle for which $f(\theta) = 1$ for $0 \le \theta \le \thetadyn$ is given
by $\thetadyn = \tan^{-1}(\Rcy/\Zcy)$. The procedure for determining the
geometry of the H~II region and the fraction of escaping photons is similar
to that of the superbubbles.  The only difference is the treatment of the
shell. As with the superbubble, we assume that the density of the shell is
$\nsh = \mach$$^2$ \nH\ \ and that the column density of the shell is equal
to the column density of swept-up gas within the cavity. However, for
dynamic chimneys, the trajectories of the shell deviate from radial lines
since the superbubble evolves from a sphere into a cylindrical
cavity. Technically, the column density of the shell for a vertical height
$Z$ would be given by the line integral of the density of the diffuse ISM
along the trajectory that leads to the chimney wall at a height $Z$. The
trajectories can be determined by requiring them to be perpendicular to the
surface of the expanding shock wave.  For an isothermal atmosphere,
\cite*{newman99a} found an analytic solution for the trajectories as a
function of time, and, for heights less than a scale height or so, the
particle trajectories are found to be roughly radial. For larger heights,
the trajectories tend to bend away from the vertical axis and become
horizontal near the shell boundary. Therefore, we make the approximation
that the streamlines are purely radial for $Z\le \Ztrans$ and purely
horizontal for $Z>\Ztrans$, where $\Ztrans \sim 2-3\sigmah$, a free
parameter of the model.  As discussed below, the results of this paper are
insensitive to the value of \Ztrans.  With this treatment, for $Z<\Ztrans$,
the shell thickness is given by
\begin{equation}
\nsh \frac{\Delta R(Z)}{\sin\theta} = \int\limits_{0}^{l}\nH \del R,
\end{equation}
where $l = \sqrt{\Rcy^2 + Z^2}$ and $\theta = \tan^{-1}(\Rcy/Z)$. Defining
$\Rsb = l$ and $\Rtr = l + \Delta R(Z)/\sin\theta$, the geometry of the
H~II region and the fraction of escaping radiation are again given
by equations (\ref{eq:HIIgeom}) and (\ref{eq:etaesc}), respectively.

\section{Results}
\subsection{Fraction of Escaping Photons as a Function of Time}
\subsubsection{Coeval Star Formation}
\begin{figure}
\centerline{\includegraphics[width=.5\textwidth]{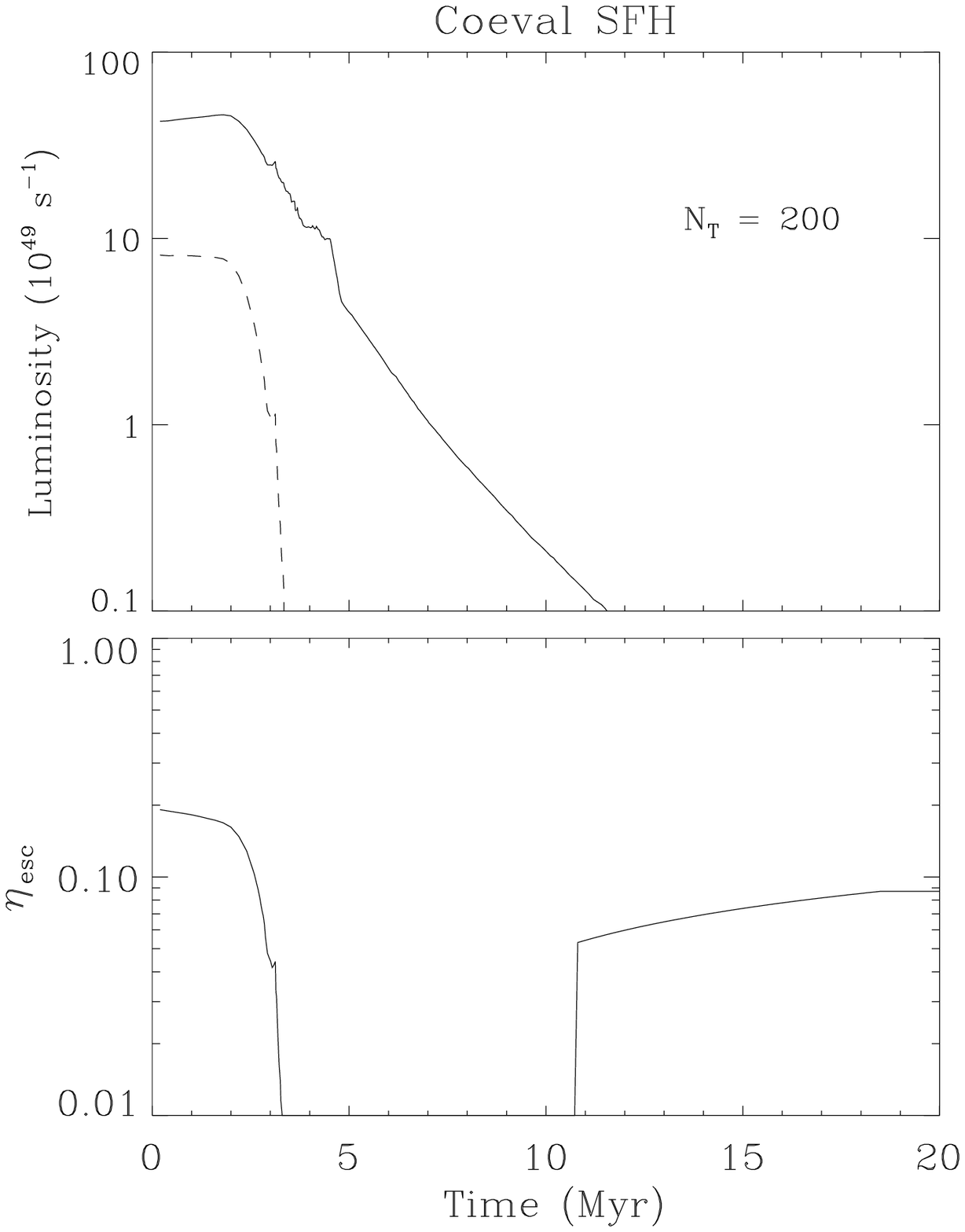}%
\hfill%
\includegraphics[width=.5\textwidth]{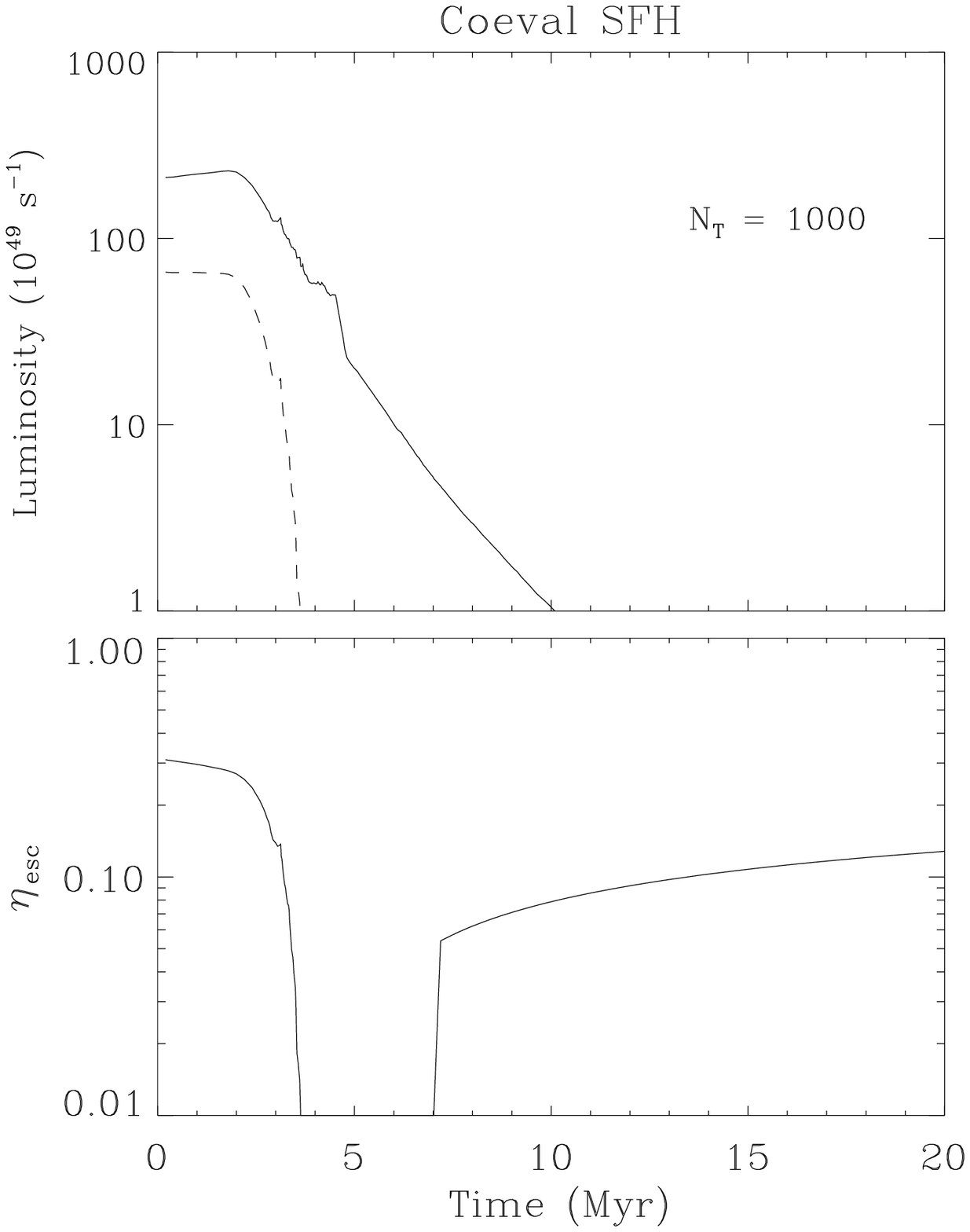}}
\figcaption{Photon luminosity emitted by a single OB association (solid line)
and the luminosity of photons escaping each side of the H~I disk (dashed
line) for coeval star-formation. Also shown is the fraction of photons
emitted that escape, $\etaesc(\NT,t) = \Sesc(t)/S(t)$. Here, $\Ztrans =
2\sigmah$, the cavity height is $\Zcy = 4\sigmah$, and $\beta =1$. The
I-front is trapped within the shell during the time interval where $\etaesc
= 0$. After blowout, $\etaesc$ suddenly increases even though the photon
luminosity of the association is relatively low at these times.
\label{fig:etaesct-coeval}}
\end{figure}

In Figure \ref{fig:etaesct-coeval} we show the photon luminosity,
emitted by a single OB association having $\NT$ stars, and the
luminosity of photons escaping each side of the H~I disk as a function
of time. Here, the stars are assumed to form coevally.
Notice that the fraction of escaping photons, $\etaesc(\NT,t)$, rapidly
decreases for $t\gta 3$ Myr. In fact, for $\NT = 1000$, roughly 90\% of the
time-integrated flux of escaping radiation occurs within the first 2.7 Myr,
while only 72\% of the time-integrated flux of ionizing radiation has been
emitted by this time. The reason for this rapid decrease is that, as the
column density of the shell accumulates, the ionizing radiation is very
quickly trapped within the shell, causing the fraction of escaping photons
to drop to zero. Once the ionization front is trapped within the shell, the
fraction of escaping photons remains zero unless the superbubble blows out
of the H~I disk. (We have not considered shell fragmentation). If blowout
occurs, photons can directly escape through the chimney (a small
fraction can also escape by penetrating through the upper H~II disk),
causing $\etaesc$ to jump up to $\sim 10$\%. However, by the time of
blowout, which occurs at $t\sim 7.0$ Myr for $\NT=1000$ and $10.0$ Myr for
$\NT=200$, the photon luminosity is considerably lower than its peak
luminosity (which occurs at $t \sim 2$ Myr). Therefore, the luminosity
of escaping radiation is significantly lower compared to its value
for $t\lta 2$ Myr.

\subsubsection{Gaussian Star Formation}
\begin{figure}
\centerline{\includegraphics[width=.5\textwidth]{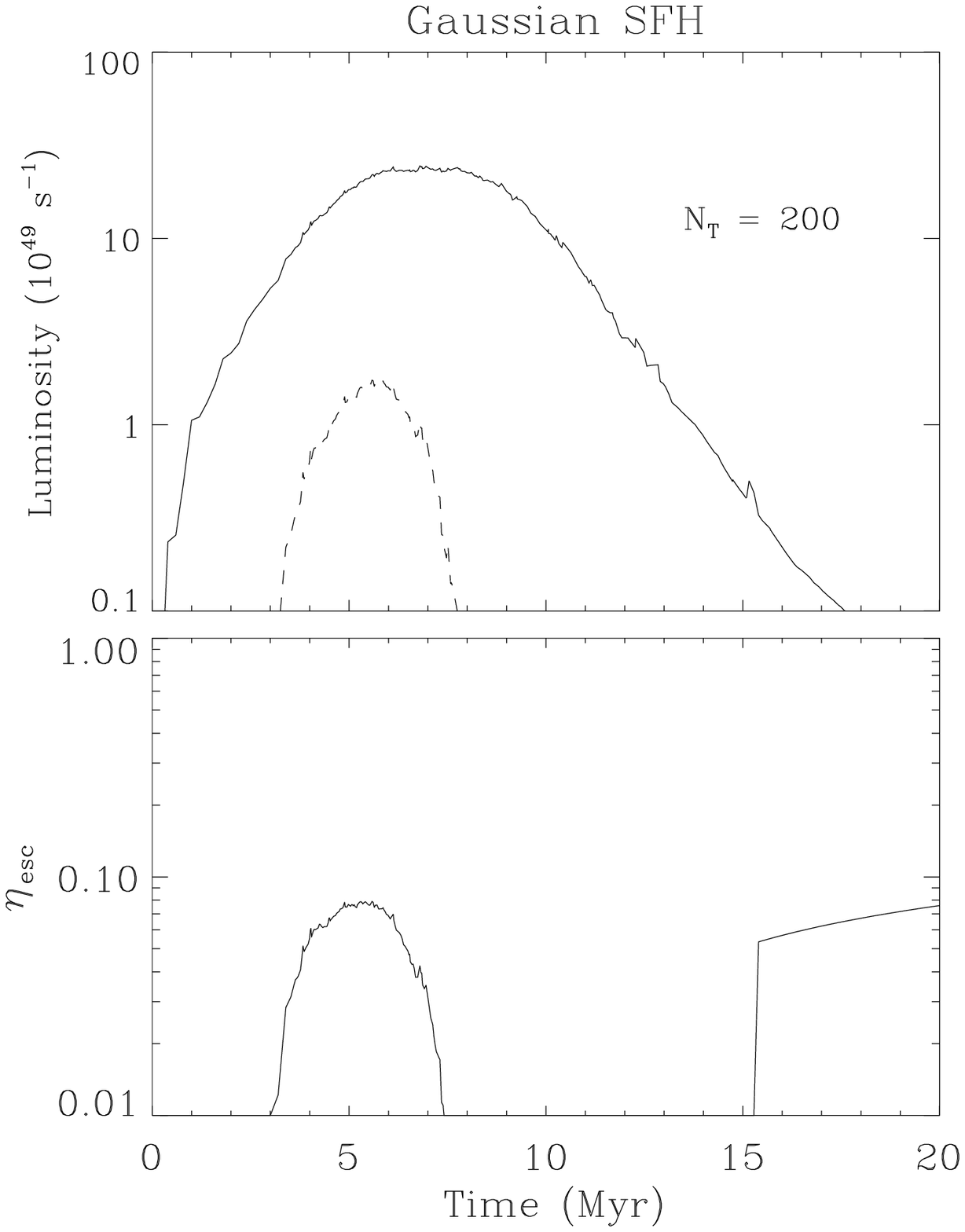}%
\hfill%
\includegraphics[width=.5\textwidth]{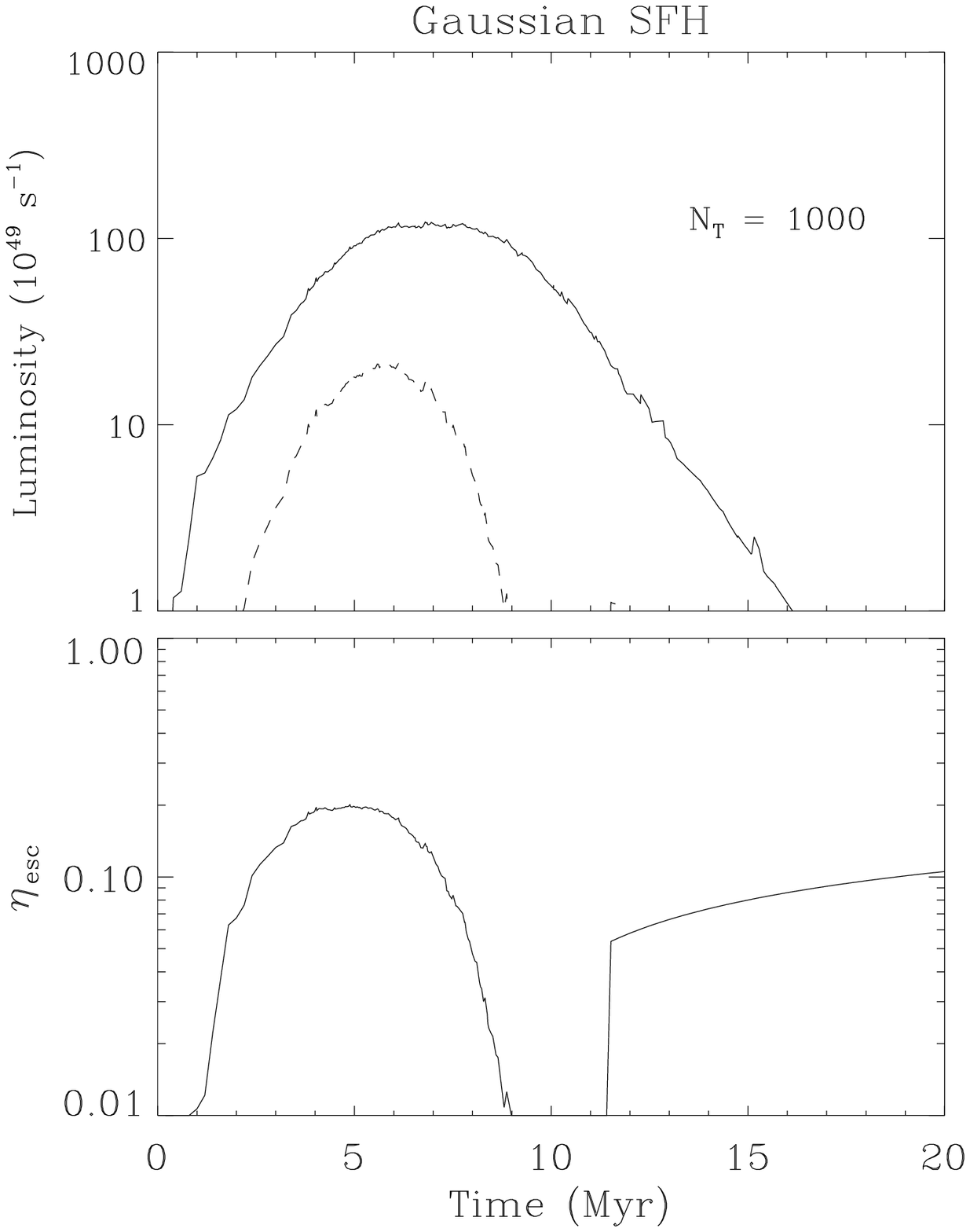}}
\figcaption{Photon luminosity emitted by a single OB association (solid line)
and the luminosity of photons escaping each side of the H~I disk (dashed
line) for a Gaussian SFH with $\sigmat = 2$ Myr. Also shown is the
fraction of emitted photons that escape, $\etaesc(\NT,t) =
\Sesc(t)/S(t)$. Here, $\Ztrans = 2\sigmah$, $\Zcy = 4\sigmah$,
and $\beta =1$.
\label{fig:etaesct-gauss}}
\end{figure}

\begin{figure}
\centerline{\includegraphics[width=.8\textwidth]{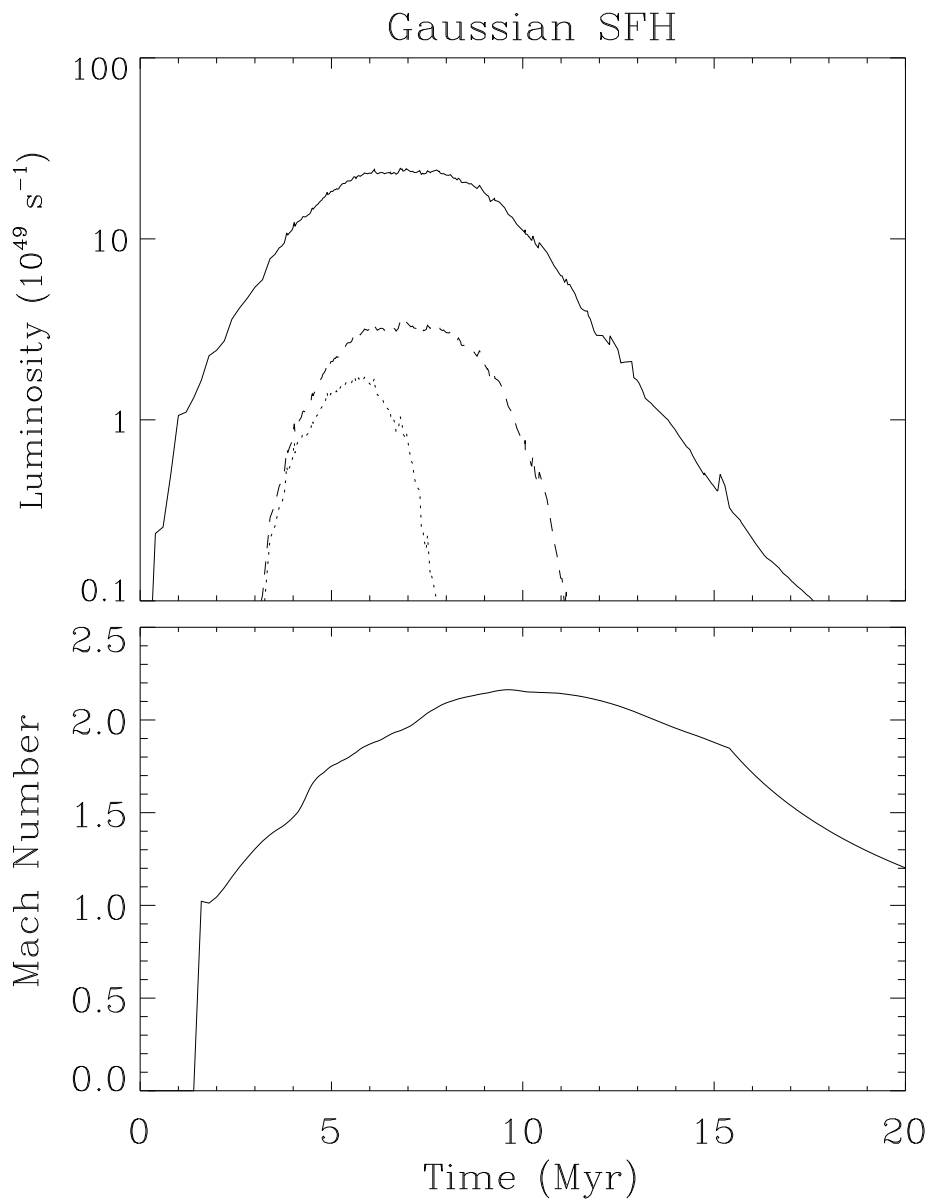}}
\figcaption{Photon luminosity emitted by a single OB association (solid line)
of size $\NT=200$ and the luminosity of photons escaping each side of the
H~I disk (dotted line) for a Gaussian SFH with $\sigmat = 2$ Myr.
Also shown is the luminosity of escaping photons for the case where the
dynamics of the bubble is not considered (dashed line).  The
bottom panel shows the Mach number of the shock front as a function of
time. Due to the high density of the evolving shell, the luminosity of
escaping radiation is reduced for $t\gta 4$ Myr.
\label{fig:etaesct2}}
\end{figure}

In Figure \ref{fig:etaesct-gauss} we plot the emitted and escaping photon
luminosity luminosity for case of the noncoeval, Gaussian SFH with $\sigmat
= 2$ Myr.
For $t=0$, the photon luminosity $S(t)$ is very low, and therefore the
fraction of escaping photons is small or even zero, in accordance with the
results of our DS94. As the luminosity
increases, $\etaesc$ increases until the shell of the superbubble becomes
thick enough that the shell-integrated recombination rate approaches the
photoionization rate. Since the shell expands more slowly compared to
the coeval SFH (having the same value of \NT), the column density of the
shell does not increase as quickly. In addition, due to the lower Mach
number of the shock front, the shell density is lower than that of the
coeval SFH. For these two reasons, \etaesc\ does not fall off in time as
quickly as in the coeval case after it reaches its peak value. Nevertheless,
the shell does attenuate significantly the flux of escaping radiation.  In
Figure \ref{fig:etaesct2}, for $\NT = 200$, we compare the flux of escaping
radiation as a function of time with that for the case where no dynamics
are considered (e.g., as done by DS94 but using the
time-dependent photon luminosity). 
For $t \gta 5$ Myr, the shell becomes becomes efficient in reducing the
escaping flux.  Therefore, as with the coeval SFH, most of the escaping
radiation escapes early, within a short time interval relative to the
lifetime of the OB association.  The time-integrated flux of photons that
escape the disk after blowout is considerably small relative to the
lifetime-integrated flux. For example, for $\NT = 1000$, fewer than 3\% of
the photons that escape the disk over the lifetime of the association
escape after blowout (while fewer than 4\% of the photons emitted by the
association are produced after blowout). For $\NT = 200$, approximately 1\%
escape after blowout. These values are not much different than for the
coeval SFH. The reason for this insensitivity is that as $\sigmat$ is
increased ($\sigmat=0$ for coeval star-formation), the blowout timescale is
also increased. Thus, even though there are more stars alive at longer
times for noncoeval SFHs, there are still relatively few alive at the time
of blowout.

\subsection{Time Integrated Fraction of Escaping Photons}\label{sec:etaescavg}
\begin{figure}
\centerline{\includegraphics[width=.8\textwidth]{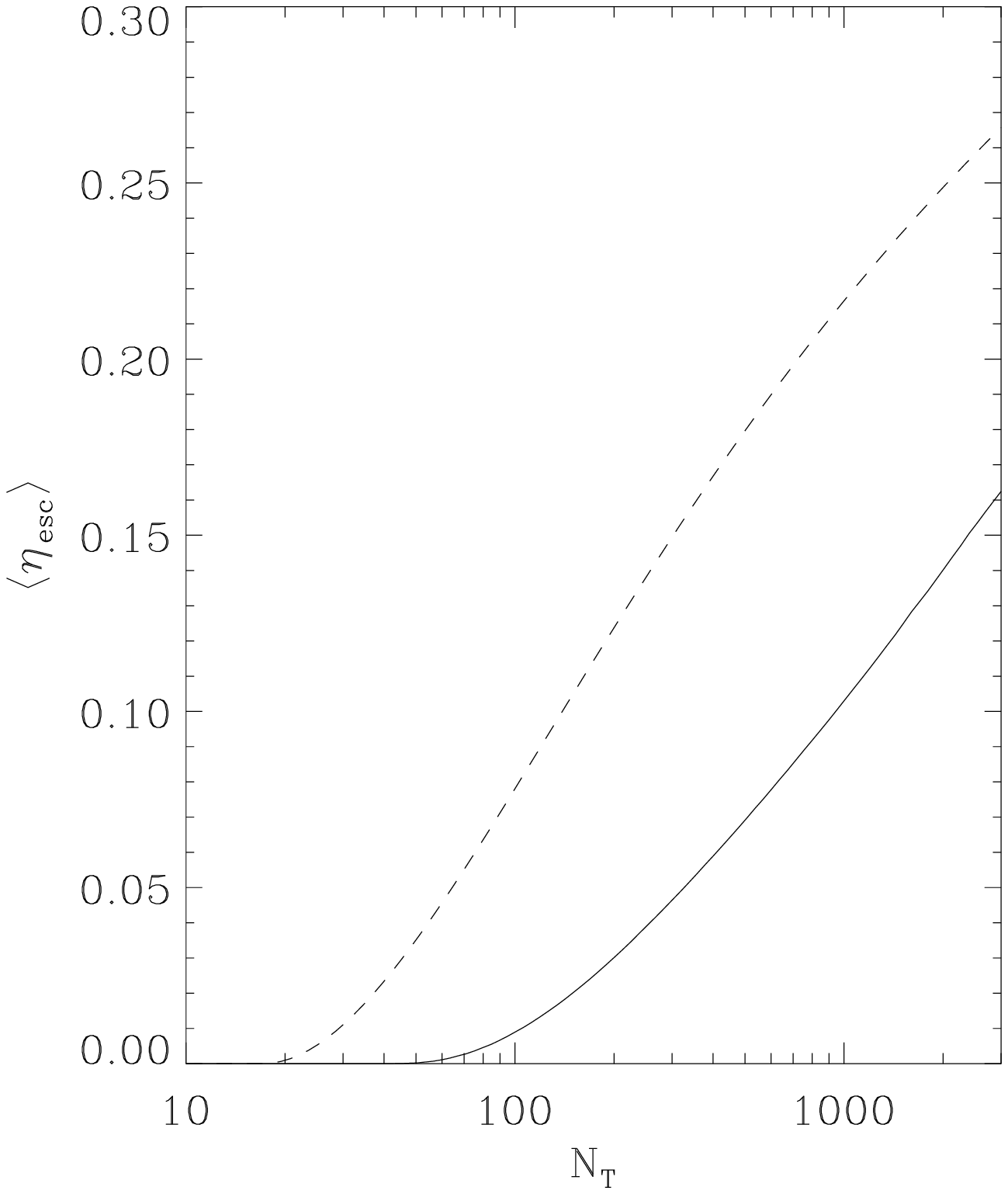}}
\figcaption{Time-integrated fraction of escaping photons as a function of
association size \NT. Here, $\Ztrans = 2\sigmah$, $\Zcy = 4\sigmah$,
and $\beta =1$. Solid line: Gaussian SFH, $\sigmat = 2$
Myr.  Dashed line: Coeval SFH.
\label{fig:time-int-eta}}
\end{figure}

The time-integrated fraction of escaping photons for each side of the H~I
disk by an association of size $\NT$ is given by
\begin{equation}
\avg{\etaesc(\NT)} = \frac{\int\limits_{0}^{\infty} \etaesc(\NT,t)
S(\NT,t)\ \del t}{\int\limits_{0}^{\infty}S(\NT,t)\ \del t}.
\end{equation}
In Figure \ref{fig:time-int-eta} we plot $\avg{\etaesc(\NT)}$ for both the
coeval SFH and the noncoeval Gaussian star-formation
model. For reference, the maximum photon luminosity of an association over
its lifetime is $\Snot(\NT) = 0.231\NT$ ($10^{49}$ s$^{-1}$) for
coeval star-formation and $\Snot(\NT) = 0.122\NT$ ($10^{49}$
s$^{-1}$) for the Gaussian SFH.
Note that $\avg{\etaesc}$ is smaller for all values of \NT\ for the Gaussian
SFH. There are several reasons for this difference. One
reason is that, for a given value of \NT, the Gaussian SFH
has a peak photon luminosity that is roughly a factor of two smaller than
that of the coeval model. Therefore, even if the shell structures of the
superbubble were identical during the interval of maximum luminosity, the
maximum value of \etaesc\ for the Gaussian model would be smaller since
$\etaesc$ increases with increasing photon luminosity (DS94). In
addition, the luminosity profile $S(t)$ for the coeval model is much more
peaked, with $\sim 70$\% of its lifetime-integrated luminosity
emitted within the first $2.5$ Myr,
compared to the relatively broad distribution of $S(t)$ for the Gaussian
SFH.  Thus, since $\avg{\etaesc}$ is weighted by $S(t)$,
and $\etaesc(\NT)$ is near its peak value while $S(t)$ is near its peak
value, the coeval SFHs yield higher values. Finally, for
coeval SFHs, the peak of $S(t)$ occurs shortly after
the formation of the association, before the formation of a significant
superbubble shell, allowing these photons to escape more easily.

\subsection{Fraction of Escaping Ionizing Photons}
We now discuss the fraction of photons, currently being
emitted by OB associations of all ages and sizes, that escape
the H~I disk of the Galaxy. 

\subsubsection{Luminosity Function}
We assume a constant formation rate, \dotNass, of associations per unit
area in our Galaxy. The probability of a new association having a peak
luminosity \Snot\ [recall that $S(\NT,t) = \Snot(\NT) y(t)$] is given by
the implicit association luminosity function, $\del N/\del\Snot$.
Therefore, during a small time interval $\delta t$, the number of
associations that will eventually have a peak luminosity \Snot\ is $\delta
N(\Snot)= \dotNass \del N/\del\Snot\ \delta t$.  The number density
(kpc$^{-2}$ s$^{-1}$) of associations observed today as a function of
luminosity, $\del\Nass(S)/\del S$, is related to the implicit luminosity
function via
\begin{equation}\label{eq:dNassdS}
\frac{\del \Nass(S)}{\del S} = \int\limits_{\tmin(S)}^{\tmax(S)} \dotNass
\frac{\del N}{\del\Snot}\left[\Snot=S/y(t)\right]\ \del t.
\end{equation}
Here, $\tmax(S)$ is given by $\tmax = y^{-1}(S/S_2)$, where $y^{-1}$ is the
inverse function of $y(t)$, and \tmin\ is given by
\begin{equation}
\tmin = \left\{\begin{array}{r@{\quad:\quad}l}
               0             & S \ge S_1\\
               y^{-1}(S/S_1) & S <   S_1. \end{array}\right. 
\end{equation}
If the dimming function $y(t)$ is multi-valued, then $\tmax$ and $\tmin$
(for $S < S_1$) are determined from its maximum values.

The observed luminosity function, from H$\alpha$ measurements of all known
H~II regions in 30 nearby spiral and irregular galaxies, is given by
$\del\Nass(S)/\del S \propto S^{-\Gamma}$ for $S_1 \le S \le S_2$, where
$\Gamma =2.0 \pm 0.5$ (\cite{kennicutt89a,banfi93a,rozas96a}, and
references therein).  \cite*{mckee97a} finds a similar result from studying
Galactic H~II regions. See DS94 for a discussion regarding the uncertainty
of the upper and lower limits of the luminosity function.  Assuming $\del
N/\del\Snot \propto \Snot^{-\Gamma}$ with $\Gamma \gta 1$, the
corresponding $\del\Nass(S)/\del S$ is close to a pure power law with the
same power-law index, with deviations first occurring at $S \sim
S_2$. Therefore, for the remainder of this paper, unless otherwise stated,
we assume that the intrinsic luminosity function is a power law with
$\Gamma = 2$.

A better understanding of the relationship between the implicit luminosity
function and the observed luminosity function can be obtained by
considering an example with a simple analytical dimming function,
$y(t)$. Assume $y(t) = \exp(-t/\tau)$ and the implicit luminosity function
$\del N/\del\Snot = N_0 \Snot^{-\Gamma}$.  Then, for $S>S_1$,
\begin{eqnarray}
\frac{\del \Nass(S)}{\del S} &=& \int\limits_{\tmin(S)}^{\tmax(S)} \dotNass
\frac{\del N}{\del\Snot}(\Snot=S/y(t))\ \del t \\
&=&\dotNass N_0\int\limits_{0}^{\tmax}\left(\frac{S}{y(t)}\right)^{-\Gamma}
\del t \nonumber \\ 
&=&\dotNass N_0 S^{-\Gamma} \int\limits_{0}^{\tmax}\exp(-\Gamma
t/\tau)\ \del t \nonumber \\
&\propto& S^{-\Gamma}[1-\exp(-\Gamma\tmax/\tau)].
\end{eqnarray}
Here, $\tmax = y^{-1}(S/S_2) = -\tau\ln(S/S_2)$, so finally we have
\begin{equation}
\frac{\del\Nass(S)}{\del S} \propto
S^{-\Gamma}\left[1-\left(\frac{S}{S_2}\right)^{\Gamma}\right].
\end{equation}
Therefore for $S>S_1$ and $S\ll S_2$, $\del\Nass/\del S \propto
S^{-\Gamma}$. The larger the value of $\Gamma$, the closer $S$ must be to
$S_2$ for appreciable differences to occur between the intrinsic and
observed luminosity functions. For $\Gamma =2$, a 10\% difference between
the two functions first occurs when $S/S_2 > 0.32$. Since the range of
association luminosities is several orders of magnitude, and the value of
$S_2$ is very uncertain, we regard this discrepancy as negligible. Note
that the reason for any differences between the two luminosity functions is
due to our prescription that the implicit luminosity function is truncated
at $S_2$. Physically, as $S$ approaches $S_2$, only the largest
associations recently formed can contribute to the luminosity function,
causing the observed luminosity function to drop to zero as $S$ reaches
$S_2$. Also note that the difference between the two luminosity functions
is independent of the functional form of the dimming function.

\subsubsection{Production Rate and Escaping Rate of Ionizing Photons}
\begin{figure}
\centerline{\includegraphics[width=.8\textwidth]{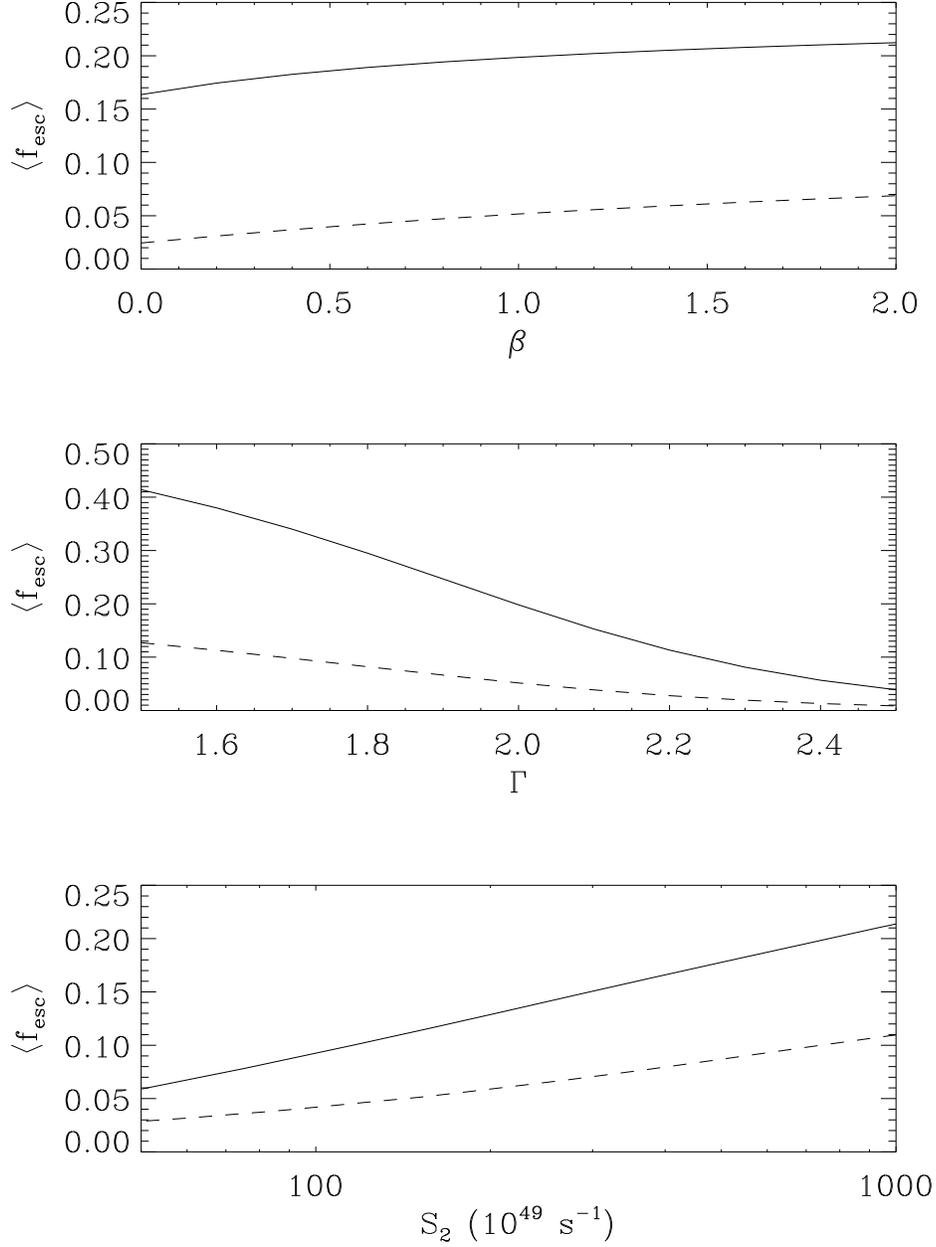}}
\figcaption{Fraction of Lyc photons, currently being emitted, that escape each
side of the H~I disk. Solid line: coeval SFH.  Dashed line: Gaussian SFH
with $\sigmat = 2$ Myr. Top panel: $\Gamma = 2$ ($\del N/\del S_0 \propto
S_0^{-\Gamma}$) and the maximum association size is $\NT = 3200$
(corresponding to $S_2 = 7.4\times 10^{51}$ s$^{-1}$ for the coeval star
formation history and $S_2 = 3.9\times 10^{51}$ s$^{-1}$ for the Gaussian
SFH).  Middle panel: $\beta = 1$, and the range of association sizes is the
same as given above.  Bottom panel: $\Gamma = 2$ and $\beta=1$. For all
models, $\Ztrans = 2\sigmah$.
\label{fig:fesc-vs-parms}}
\end{figure}

The current production rate of ionizing photons (\# per unit area per unit
time) is given by
\begin{equation}
\Psilyc = \int\limits_{0}^{S_2} S \frac{\del\Nass}{\del S}\ \del S.
\end{equation}
Substituting the expression for the observed luminosity function in
terms of the implicit luminosity function [equation (\ref{eq:dNassdS})], we find,
\begin{equation}
\Psilyc = \dotNass\int\limits_{0}^{S_2}S\int\limits_{\tmin}^{\tmax}\frac{\del
N}{\del\Snot}(\Snot = S/y(t))\ \del t\ \del S.
\end{equation}
The flux of photons escaping from each side of the H~I disk is given by
\begin{equation}\label{eq:Psiesc}
\Psiesc = \dotNass\int\limits_{0}^{S_2}S\int\limits_{\tmin}^{\tmax}
\etaesc(S,t)\frac{\del
N}{\del\Snot}\left[\Snot = S/y(t)\right]\ \del t\ \del S.
\end{equation}
Here, $\etaesc(S,t)$ is the fraction of photons that escape each side of
the H~I disk emitted by an association of size $\NT$, where the peak
luminosity of this association is $\max[S(\NT,t)] = \Snot(\NT)$. By
computing a two-dimensional grid of $\etaesc(\NT,t)$ using the methods
outlined in \S\ref{sec:etaesc}, we integrate equation (\ref{eq:Psiesc}) for
the two star-formation scenarios. For given values of $S$ and $t$, an
interpolation routine was used to find the corresponding value of $\NT$,
such that $\etaesc(S,t) = \etaesc[\NT(S,t),t]$.  Finally, the total
fraction of photons, currently being emitted, that escape both sides of
the H~I disk is
\begin{equation}
\fesc = 2\frac{\Psiesc}{\Psilyc}.
\end{equation}
We find that $\fesc \sim 12$\% for the coeval SFH and $\fesc \sim 6$\% for
the Gaussian SFH.  In Figure \ref{fig:fesc-vs-parms}, we show how \fesc\
depends on the model parameters.  For reasons given in
\S\ref{sec:etaescavg}, \fesc\ for the Gaussian SFH is considerably lower
than that for the coeval SFH. By far, \fesc\ is most sensitive to the
power-law index of the luminosity function, $\Gamma$. This sensitivity
arises because most of the escaping radiation comes from the relatively few
large associations with $\NT \gta 100$; small changes in $\Gamma$ cause
rather large changes in the relative number of these large associations.
For the coeval SFH, \fesc\ is insensitive to the magnetic field (the
parameter $\beta$).  As the superbubble expands, the column density of the
shell increases quickly, causing $\etaesc(S,t)$ to drop to zero very
quickly even for strong magnetic fields. For the Gaussian SFH, the
superbubble expands more slowly, and different values of $\beta$ cause the
shell to trap the ionization front at different times. Since this trapping
of the I-front occurs near the peak of $S(t)$, \fesc\ is more sensitive to
$\beta$. Lastly, we do not plot how \fesc\ depends on the parameters
involved with the superbubble blowout (e.g., \Ztr\ and \Zcy), as we found
that varying these parameter over a range between $\sigmah$ and $4\sigmah$
corresponded to a relative change of \fesc\ less than 1\%.  This
insensitivity occurs because most emitted and escaping photons occur early
after the formation of the association.

\section{Complexities to the Model}

We now discuss the key issues not included in our standard model. We then
discuss how these effects may alter our results.

\subsection{Shell Instabilities}\label{sec:instabilities}

As discussed above, the presence of a thick neutral shell of cool, shocked
ISM produced by multiple supernova explosions is quite effective in
absorbing the radiation coming from the OB association. This process
significantly reduces the fraction of photons able to escape into the
galactic halo until blowout takes place. At that stage, the previously
decelerating shell is suddenly accelerated and the interface between the
dense shell and the hot cavity gas becomes Rayleigh-Taylor (R-T)
unstable. The nonlinear development of such instability induces
fragmentation of the shell into a number of clumps, through which the
photons can percolate and almost freely escape into the halo.  This process
generally occurs relatively late (or not at all), when the radiative flux
has already decreased by a considerable amount due to the aging of the
parent OB association. It is thus worthwhile to investigate if the
fragmentation process driven by R-T or other instabilities can take place
before blowout, consequently increasing \fesc.

There are at least three types of instabilities that could affect the shell
at the beginning or during the pressure-driven radiative (PDR) phase: 1)
dynamical, 2) Rayleigh-Taylor and 3) gravitational.  To
assess the importance of these processes, it is necessary to discuss in more
detail the development of the shell structure for SN bubbles. We restrict
our discussion to uniform ambient gas density, which is a reasonable
approximation to study the pre-blowout phase.  The structure of a
wind-blown bubble (the energy injection by multiple SNe can be well
approximated as a continuous process) is constituted by a reverse (``wind")
shock propagating back into the wind, which is under most conditions
adiabatic, and an ``ambient" shock propagating into the surrounding medium;
the two shocks are separated by a contact discontinuity. Initially, the
ambient shock is also adiabatic and its radius increases with time as
$t^{3/5}$.  However, when the cooling time of the shocked ambient gas
becomes of the order of the age of the bubble, the ambient shock becomes
radiative and the shocked ambient gas collapses into a thin shell.

The shell-formation phase is characterized by violent dynamical
instabilities which tend to amplify the perturbation due to classical
thermal instability in the catastrophically cooling shocked ambient gas.
Are these instabilities able to disrupt the shell soon after its formation?
At least three different dynamical instabilities may arise in the thin,
cool shell: nonradial oscillations of the radiative shock
(\cite{bertschinger86a}), the nonlinear thin-shell (NTSI) instability
arising when the shell is bounded by two radiative shocks
(\cite{vishniac83a}), and the pressure-driven thin-shell overstability
(PDTSO) found in a shell bounded by a high thermal pressure on one side and
by a radiative shock on the other (\cite{vishniac94a}). These dynamical
instabilities may produce considerable distortions in the shell, often
showing nonradial oscillations. The main difference between the latter two
instabilities is that while the NTSI grows exponentially, the PDTSO only
grows with a power-law rate. Recent numerical simulations
(\cite{maclow93a,blondin98a}) have shown that the PDTSO is probably the
most dangerous instability, especially for superbubbles where the wind
shock is adiabatic rather than radiative for a large fraction of its
evolution.  However, the PDTSO works only for the time during which the
Mach number of the shell is larger than $\sim 3$.  As the PDTSO grows
with time as $t^s$ with $s\approx 1/2$, whereas the shock velocity
decreases as $t^{-2/5}$ (see equation [\ref{radius}] below) in the
radiative phase, amplifications $\simlt 30$ are typically possible,
i.e. the overstability saturates.  The above numerical studies concluded on
this basis that fragmentation of the shell due to these instabilities is
unlikely, although \cite*{blondin98a} emphasize that in a denser ISM the
shock enters the radiative phase at a higher Mach number, thus allowing for
a longer growth time of the instability.

Even if the shell can survive dynamical instabilities, it can be fragmented
by R-T instabilities generated {\it before} blowout.  As the radiative
bubble expands, its pressure drops and eventually becomes equal to the that
of the ISM.  In this case, the wind shock stalls, but the shocked wind is
still adiabatic and forces the shell to continue its expansion.  However,
the expansion rate is now so slow that the acceleration of the shell is
dominated by the galactic gravitational field. Thus, the polar region of
the shell will be R-T unstable even in the absence of a blowout. We
calculate the time at which the pressure-confined phase occurs (and hence
the R-T instability starts to grow) as follows.  For constant luminosity,
the evolution of the shell radius in the radiative phase is given by
(\cite{castor75a,weaver77a,ostriker88a})
\begin{equation}
\label{radius}
R(t) \simeq 66 \left({L_{38}\over n_0}\right)^{1/5} t_6^{3/5} {\rm ~~pc}
\end{equation}
where $L_{38}\equiv L/(10^{38} {\rm ~erg~s}^{-1})$ is the wind mechanical
luminosity, $n_0$ is the ambient gas number density, and $t_6\equiv t/(10^6
{\rm ~yr})$. From the pressure balance condition $\dot R^2 = c_s^2$,
where $c_s$ is the ISM effective sound speed, this transition 
occurs at
\begin{equation}
\label{tpres}
t_p \simeq 3\times 10^7 \left({L_{38}\over n_0 c_{s,10}^5}\right)^{1/2} {\rm ~~yr}
\end{equation}
where $c_{s,10}\equiv c_s/(10\ {\rm km\ s}^{-1})$ is the ISM effective sound
speed.  The growth time of the R-T instability on spatial scales $\lambda_s
\sim \sigmah$ is
\begin{equation}
\label{RTgrow} 
t_{\rm RT} \simeq \left[{{2\pi g \over \lambda_s } {(\Delta - 1)\over (\Delta +
1)}}\right]^{-1/2} \simeq \left({2\pi g \over \lambda_s }\right)^{-1/2}
\sim
1{\rm ~Myr}
\end{equation}
if the density contrast between the shell and the hot cavity gas $\Delta
\equiv n_s/n_h \gg 1$ and $\avg{g} = 10^{-8}$ cm s$^{-2}$. As $t_{\rm RT}$
is much shorter than $t_p$, the shell is fragmented soon after it enters
the pressure-confined phase.  For the case of a Gaussian SFH, the
mechanical luminosity is initially very low, causing the shock to stagnate
much earlier than the constant-luminosity case.

In addition, the shell can be re-accelerated early in its evolutionary
stage shortly after the first supernovae of the association due to the
sudden increase in the mechanical luminosity. We have numerically evaluated
$\avg{g}$ due to this process assuming the Kompaneets approximation is
valid, and find that the product of $\avg{g}$ and the duration of the
re-acceleration (typically $\sim 1-4$ Myr) is $\sim 5$.  Thus, the growth
of the instability is moderately non-linear.  A more detailed study of this
process, including the role of poisoning, will be given in \cite*{dove99b}.
Therefore, in contrast to the case of a constant mechanical luminosity, the
shell can be R-T unstable before blowout. In this case, a fragmenting shell
would allow a higher amount of ionizing radiation to escape the H~I disk
since fragmentation occurs during the time interval in which the photon
luminosity of the association is near its peak value.

The shell might also be fragmented by a gravitational instability. Several
studies have investigated this possibility, both under interstellar
(\cite{elmegreen77a,mccray87a,voit88a}) and intergalactic conditions
(\cite{ostriker81a,couchman86a}). Here, we simply state the the growth time
of the most unstable mode is
\begin{equation}
\label{ggrow} 
t_{G} \simeq {3 c_{s,shell}^2\over \pi G \rho_0 R(t)}=
2\times 10^7 L_{38}^{1/8} n_0^{-1/2}{\rm ~yr}. 
\end{equation}
Again, we see that this mechanism is able to fragment the shell on time
scales of interest and similar to the ones found for the R-T instability.
We conclude that, even in the absence of blowout, fragmentation of the
shell is likely about $3\times 10^7$~yr after the beginning of the energy
injection.

Even if the shells of superbubbles do not blow out, do they fragment while
the OB association is still producing ionizing photons?  Simple conditions
on the occurrence of blowout have been derived in different contexts by
\cite*{maclow88a} and \cite*{maclow99a}. Neglecting the role of magnetic
fields, \cite*{koo92a} studied in detail the case for the Milky Way and
concluded that the blowout requires $L_{38} \simgt 4$. For a marginally
confined bubble, we find from equation (\ref{tpres}) and equation
(\ref{ggrow}) that $t_p \approx 6\times 10^7$~yr and $t_G \approx 2.4\times
10^7$~yr for the R-T and the gravitational instability, respectively.  Note
that, as $L_{38}$ is decreased, fragmentation occurs earlier in time.
Thus, even bubbles that do not blow out of the disk can fragment on time scales
short enough that at least some ionizing photons escape. Determining the
covering fraction of the fragmented clumps is outside the scope of the
present estimates and will require numerical simulations of the
fragmentation process.  As a final caveat, we stress that a dynamically
important magnetic field could qualitatively modify the above conclusions
by suppressing some of the instabilities (\cite{tomisaka98a}) and by
introducing new ones.

\subsection{Clumpiness of the Diffuse H~I Gas}
Although our models assume that, outside the superbubbles, the neutral gas is
distributed diffusely, there is observational evidence that a significant
fraction of the gas is in cold H~I clouds, with densities much larger than
the space-averaged density. Due to the higher recombination rates, these
clouds modify our radiation-transfer results and therefore merit some
consideration. Since most radiation that escapes the H~I disk comes from
young OB associations, with shells having small column densities, we
considered the ramifications of a clumpy medium by ignoring the effects of
evolving superbubbles. A more proper treatment would consider the
time-dependent evolution of the clouds due to passage of the shell of the
superbubble as well as the expansion and evaporation of the
clouds after being photoionized.

As an approximate model, we let a mass-fraction of the gas, \acl, be
contained in cold, dense clouds. The remaining fraction of the gas, $(1-\acl)$, is
distributed diffusely using the Dickey-Lockman density distribution (Dickey
\& Lockman 1990).  We consider two cloud geometries: (1) spheres of
radius $\rc$; or (2) cylindrical disks with radius $\rc$ and an aspect ratio
$\deltap = h/\rc$, where $h<\rc$ is the height of the cylinder.  Each of
the cylindrical disks is assumed to be ``face-on'' with respect to the line
of sight from the OB association. For either geometry, all clouds are
assumed to have identical column densities, sizes, and densities.  The
number of clouds per unit area of the disk is given by
\begin{equation}
\Nc = \frac{\acl N_{HI}}{\Vc\ \ncl},
\end{equation}
where $\Vc$ is the volume of each cloud, $\ncl$ is the density of hydrogen
within each cloud, and $N_{HI} = 6 \times 10^{20}$ cm$^{-2}$ is the
observed total column density of hydrogen through the entire disk at the
solar circle.   

We assume that the clouds are distributed randomly within the disk of the
galaxy using a probability distribution, $P_{cl}(Z)$, that is constant in
the plane and proportional to the Dickey-Lockman density distribution in
the vertical direction.  The {\em average} number of clouds intersecting a
line of sight having an inclination angle $\theta$ ($\mu = \cos\theta$),
emanating from an OB association situated on the midplane, is given by
\begin{equation}\label{eq:avgclouds}
N_{int} = \frac{\pi \rc^2 \Nc}{2\mu} = \left\{\begin{array}{l@{\quad:\quad}l}
\frac{1.6}{\mu} \left(\frac{\acl}{0.5}\right)\ \left(\frac{\rc}{10\ {\rm
pc}}\right)^{-1}\ \left(\frac{\ncl}{30\ {\rm cm}^{-3}}\right)^{-1}\
\left(\frac{\deltap}{0.1}\right)^{-1} & {\rm for\ cylindrical\ disks} \\
\frac{0.12}{\mu} \left(\frac{\acl}{0.5}\right)\ \left(\frac{\rc}{10\ {\rm
pc}}\right)^{-1}\ \left(\frac{\ncl}{30\ {\rm cm}^{-3}}\right)^{-1}\
 & {\rm for\ spheres} 
\end{array}\right.
\end{equation}

As in \S2.3.2, the fraction of LyC photons from a single OB association
that escape each side of the disk is given by
\begin{equation}\label{eq:etaesc2}
\etaesc(S) = \int\limits_{0}^{\pi/2} \frac{f(\theta,S)}{2} \sin(\theta)\
d\theta,
\end{equation}
where
\begin{equation}\label{eq:foftheta2}
f(\theta,S) = 1 -
\frac{4\pi\alpha_H^{(2)}}{S}\int\limits_0^{\infty} n_H^2(r)\ r^2\ dr.
\end{equation}
Here, $n_H(r)$ is given by the diffuse density distribution for radii
outside any cloud and is equal to $\ncl$ for radii inside any cloud.  For a
given value of $\theta$ and $S$, we evaluate $f(\theta,S)$ by averaging
over 100 Monte-Carlo simulations. For each simulation, the number of clouds
along the line of sight is drawn from a Poisson distribution having an
average number of clouds given by equation (\ref{eq:avgclouds}).  The
distances from the origin of each of these clouds to the OB association are
picked randomly using the probability distribution $P_{cl}(Z)$. For
spherical clouds, the average chord length between the line of sight and
the cloud is $l = h = 4\rc/3$. For the cylindrical clouds, we assume that each
cloud is orientated face-on, such that $l = \deltap \rc$. The observed column
density of each cloud in the line of sight is $\nc l$. Using the average
values of $f(\theta,S)$, we determine $\etaesc(S)$ via equation
(\ref{eq:etaesc2}). Finally, as in DS94, we determine the total fraction of
ionizing radiation that escapes the disk is determined by integrating over
the luminosity function,
\begin{equation}
\fesc = 2\ \frac{\int\limits_{S_1}^{S^2} \etaesc(S)\ \frac{d\Nass}{dS}\
dS}{\int\limits_{S_1}^{S^2} \frac{d\Nass}{dS}\ dS}.
\end{equation}

\begin{deluxetable}{lcllcl} 
\tablecaption{\fesc\ for Clumpy Static ISM\label{tab:fescclump}}
\tablewidth{0pt}
\tablehead{
\colhead{Geometry} & \colhead{$\rcl$ (pc)} & \colhead{$\acl$} & \colhead{$\deltap$} & \colhead{$S_2$ ($10^{49}$ s$^{-1}$)} & \colhead{$\fesc$} \\
}
\startdata
disk            & $5$  & $0.1$ & $0.05$ & $100$ & $0.01$ \nl \tablevspace{2pt}
disk            & $5$  & $0.1$ & $0.05$ & $250$ & $0.03$ \nl \tablevspace{2pt}
disk            & $5$  & $0.1$ & $0.05$ & $500$ & $0.05$ \nl \tablevspace{2pt}

disk            & $5$  & $0.1$ & $0.10$ & $100$ & $0.03$ \nl \tablevspace{2pt}
disk            & $5$  & $0.1$ & $0.10$ & $250$ & $0.05$ \nl \tablevspace{2pt}
disk            & $5$  & $0.1$ & $0.10$ & $500$ & $0.08$ \nl \tablevspace{2pt}

disk            & $5$  & $0.3$ & $0.10$ & $100$ & $0.002$ \nl \tablevspace{2pt}
disk            & $5$  & $0.3$ & $0.10$ & $250$ & $0.006$ \nl \tablevspace{2pt}
disk            & $5$  & $0.3$ & $0.10$ & $250$ & $0.01$ \nl \tablevspace{2pt}

disk            & $10$  & $0.3$ & $0.05$ & $100$ & $0.04$ \nl \tablevspace{2pt}
disk            & $10$  & $0.3$ & $0.05$ & $250$ & $0.07$ \nl \tablevspace{2pt}
disk            & $10$  & $0.3$ & $0.05$ & $500$ & $0.09$ \nl \tablevspace{2pt}

disk            & $10$  & $0.3$ & $0.10$ & $100$ & $0.07$ \nl \tablevspace{2pt}
disk            & $10$  & $0.3$ & $0.10$ & $250$ & $0.10$ \nl \tablevspace{2pt}
disk            & $10$  & $0.3$ & $0.10$ & $500$ & $0.13$ \nl \tablevspace{2pt}

disk            & $10$  & $0.5$ & $0.10$ & $100$ & $0.05$ \nl \tablevspace{2pt}
disk            & $10$  & $0.5$ & $0.10$ & $250$ & $0.07$ \nl \tablevspace{2pt}
disk            & $10$  & $0.5$ & $0.10$ & $500$ & $0.09$ \nl \tablevspace{2pt}

sphere             & $5$  & $0.3$ & $0.05$ & $100$ & $0.15$ \nl \tablevspace{2pt}
sphere             & $5$  & $0.3$ & $0.05$ & $250$ & $0.20$ \nl \tablevspace{2pt}
sphere             & $5$  & $0.3$ & $0.05$ & $500$ & $0.23$ \nl \tablevspace{2pt}

sphere             & $10$  & $0.3$ & $0.05$ & $100$ & $0.18$ \nl \tablevspace{2pt}
sphere             & $10$  & $0.3$ & $0.05$ & $250$ & $0.23$ \nl \tablevspace{2pt}
sphere             & $10$  & $0.3$ & $0.05$ & $500$ & $0.27$ \nl \tablevspace{2pt}

sphere             & $10$  & $0.5$ & $0.1$ & $100$ & $0.23$ \nl \tablevspace{2pt}
sphere             & $10$  & $0.5$ & $0.1$ & $250$ & $0.29$ \nl \tablevspace{2pt}
sphere             & $10$  & $0.5$ & $0.1$ & $500$ & $0.33$ \nl \tablevspace{2pt}

\enddata
\end{deluxetable}

In Table \ref{tab:fescclump}, we give \fesc\ for various values of $\acl,
\rcl$, and $\deltap$.  It is apparent that \fesc\ is sensitive to these
parameters, which are poorly constrained observationally.  For a
given value of $\acl$, \fesc\ decreases as the average number of clouds
within a line of sight increases. We find that the I-front is
trapped within most clouds. Only clouds very close to the OB association
are not able to trap the I-front and become completely ionized, a
consequence of the high recombination rate within the cloud.  Essentially
all escaping radiation emanates from lines of sight that are free of clouds. For
comparison, for the case where $\acl = 0$, $\fesc = 0.12, 0.18, 0.22$ for
$S_2 (10^{49}\ {\rm s}^{-1}) = 100, 250$, and $500$, respectively (these
results are identical to those of DS94). We conclude that, if an appreciable
amount of the H~I gas is in cold clouds having a disk geometry and
column densities and hydrogen densities similar to the H~I clouds studied by
\cite*{fitzpatrick97a}, then \fesc\ can be a factor of $\sim 2-5$ smaller
than previously estimated.  For a fixed amount of gas locked in cold
clouds, spherical clouds occupy a smaller solid angle as compared to thin
cylindrical clouds, allowing more radiation to escape.  Note that, since we
did not consider superbubbles in these calculations, these numbers are only
indicative of the {\em relative} decrease of $\fesc$ when cold clouds are
included in the model.

\subsection{Poisoning}
Although the survey used for calibrating the production rate of ionizing
photons is heavily biased towards associations far away from the parent
molecular cloud, most if not all associations were born in a molecular
cloud.  Therefore, early in the lifetime of the OB association, it is
possible that a considerable amount of matter is photoablated from the
molecular cloud. This material increases the interior density of the
bubble, leading to radiative cooling, potentially at a rate equal to the
the injected mechanical luminosity, a process independently dubbed
``poisoning'' by \cite*{mckee86a} and \cite*{shull95a}. This poisoning
process can alter the evolution of a young superbubble, as well as the
propagation of the I-front through the cavity.

To date, the photoablation rate has not been calculated 
for a time-varying luminosity of an OB cluster
with a proper radiation spectrum. In
addition, for a given mass-injection rate, solving the evolution of the injected
gas within the cavity is a complex problem. These complexities lead to
uncertainties in the radiative cooling rate, both in magnitude and in its
time evolution. Therefore, in order to estimate the degree to which 
poisoning alters the evolution of the superbubble and \fesc,
we introduced a factor, $g$, representing the fraction of the
mechanical luminosity injected into the cavity that is drained due to
radiative cooling from both photoablated gas and evaporated gas from the
dense shell.  Thus, $(1-g)\Lmech$ is the portion of energy available to do
$P dV$ work.  In what we consider an extreme case, we let $g=0.5$ for the
duration in which $R_{sh}$ is less than $30$ pc [see \cite*{shull95a} for a
discussion of when cooling due to poisoning becomes negligible].  In this
case, for our non-coeval model, we find that \fesc\ {\em increases} by
roughly 35\%. This increase occurs because the superbubble shell
expands more slowly, and the density of the shell is smaller
(it is proportional to the Mach number squared of the shell), and because the column
density of the shell increases more slowly.

On the other hand, it is possible that enough material is photoablated
into the hot cavity that, after recombining, it contributes to the
absorption of ionizing radiation emitted by the OB association,
partially offsetting the increase in \fesc\ due to the slower shell expansion.  
However, to model this process requires a knowledge of the relative 
configurations of the molecular cloud, the OB association, and the disk 
of the Galaxy. Also, the evolution of the injected gas, especially the
expansion of the gas as it expands from the molecular cloud or shell
boundaries and into the the line-of-sight between the OB association and
the halo, is a numerical problem beyond the scope of this paper.

\subsection{Triggered Star Formation}
The fraction of ionizing radiation escaping the H~I disk would be
increased if an appreciable number of OB associations were formed within 
pre-existing dynamic chimneys.  The two-dimensional filling factor of
dynamic chimneys is estimated by
\begin{equation}
f_{\rm fill} \sim f_{\rm as} \dotNass \tau_{\rm ca} \pi \Rcy^2,
\end{equation}
where $f_{\rm as}$ is the fraction of OB associations that are large
enough to produce a dynamic chimney, $\tau_{\rm ca}$ is the lifetime of the
cavity, and $\Rcy$ is the radius of the cavity. As discussed in
\S\ref{sec:instabilities}, associations having an average mechanical
luminosity $\Lmech \gta 4 \times 10^{38}$ erg s$^{-1}$ are thought to be
capable of blowing out of the Galactic disk (\cite{maclow99a,koo92a}). This
corresponds to an OB association size of $\NT \gta 400$, or a peak
luminosity $\Snot \gta S_{\rm cr} = 5 \times 10^{50}$ s$^{-1}$. The
fraction of associations that produce a dynamic chimney is then
\begin{equation}
f_{\rm as} = \frac{\int\limits_{S_{\rm cr}}^{S_2} \del N/\del
\Snot}{\int\limits_{S_1}^{S_2} \del N/\del\Snot} =
\frac{\left(\frac{S_2}{S_{\rm cr}}\right)^{\Gamma-1} -
1}{\left(\frac{S_2}{S_1}\right)^{\Gamma-1} - 1} \approx
\left(\frac{S_1}{S_{\rm cr}}\right)^{\Gamma-1},
\end{equation}
where we assumed $S_2 \gg S_{\rm cr} \gg S_1$. For $\Gamma = 2$ and $S_1 =
1\times 10^{48}$ s$^{-1}$, $f_{\rm as} \approx 2\times 10^{-3}$.  The
lifetime of the dynamic chimney is estimated to be the sound crossing time,
\begin{equation}
\tau_{\rm ch} \sim \frac{\Rcy}{c_{\rm s}} \sim 35\left(\frac{10\ {\rm km\
s}^{-1}}{c_{\rm s}}\right)\left(\frac{\sigmah}{0.18\ {\rm kpc}}\right)\ {\rm Myr}.
\end{equation}
Therefore,
the probability that an OB association is born within a dynamic chimney is
given by 
\begin{equation}
f_{\rm fill} \sim 0.18 \left(\frac{\dotNass}{6.4\ {\rm Myr\
kpc}^{-2}}\right)\left(\frac{\tau_{\rm as}}{35\ {\rm
Myr}}\right)\left(\frac{\Rcy}{2\sigmah}\right)^2,
\end{equation}
and it is unlikely that $\fesc$ is enhanced significantly by this process if
OB associations are born randomly in space. However, as suggested
by \cite{mccray87a}, the birth of OB associations could be triggered by
the passage of the shock wave associated with another superbubble. Indeed,
recent far ultraviolet, H$\alpha$, and H~I observations of several
dwarf galaxies show evidence of secondary star-formation sites on the dense
H~I rims of superbubbles (\cite{voit88a,stewart98a}).

\section{Discussion and Conclusions}
We find that $\fesc \sim 12-20$\% for the coeval star-formation history and
$\fesc \sim 4-10$\% for the Gaussian star-formation history.  For $S_2 =
2\times 10^{51}$ s$^{-1}$, $\beta = 1$, and $\Gamma =2$, we obtain
$\fesc = 13$\% for the coeval SFH and $\fesc = 6$\% for the Gaussian SFH.  
Adopting a production rate \footnote{This number is 50\% higher than the value 
used in DS94, due to updated stellar atmosphere models whose yield of ionizing
photons from OB associations is 50\% higher than previously estimated
(\cite{sutherland99a}).}, $\Psilyc = 4.95 \times 10^7$ cm$^{-2}$ s$^{-1}$,
the corresponding photon fluxes are $\Phi_{\rm LyC} = 2.1\times 10^6$
cm$^{-2}$ s$^{-1}$ for the coeval SFH and $\Phi_{\rm LyC} = 9.9\times 10^5$
cm$^{-2}$ s$^{-1}$ for the Gaussian SFH.  For comparison, $\fesc = 15$\%
using the formalism of DS94, where the photon luminosity was assumed
constant in time and dynamically evolving superbubbles were not considered.
In order to determine the degree to which the differences between our new
results and our old results are due to the treatment of expanding
superbubbles and dynamic chimneys, we also calculated $\fesc$ for the case
where we neglect entirely the superbubble dynamics but include the
time-varying photon luminosities. For the same parameters as given above,
we find $\fesc = 14$\% for the Gaussian SFH and $\fesc=16$\% for the coeval
SFH, values similar to the old model. These small differences are due
solely to the new treatment of the luminosity function of associations with
time-varying photon luminosities, as discussed in \S 3.3.1. Therefore, the
inclusion of the evolution of superbubbles {\em reduces} the fraction of
escaping radiation, as the superbubble shells trap more radiation.  For
coeval SFHs, the inclusion of the dynamics only changes \fesc\ from $8.2$\%
to $6.5$\%, since most radiation is emitted and escapes before a
substantial shell has formed. On the other hand, for the Gaussian SFH,
\fesc\ drops from $14$\% to $6$\%. This reduction of $\fesc$ occurs because
the photon luminosity peaks after a shell has formed, causing fewer photons
to escape compared to the case when superbubbles are neglected.

In DS94, we argued that \fesc\ must be $\sim 14$\% in order for the
Reynolds layer to be sustained by radiation from OB associations.  However,
using COSTAR model atmospheres, \cite*{sutherland99a} found that the yield
of ionizing photons from OB associations is 50\% higher than previously
estimated, and the production rate of ionizing photons by OB associations
in the solar vicinity is $\Phi_{\rm LyC} = 4.95 \times 10^7$ cm$^{-2}$
s$^{-1}$.  Thus, the required value of \fesc\ to sustain the Reynolds layer
is reduced to $\fesc \approx 8$\%.  The flux of escaping radiation for the
Gaussian SFH is about 25\% smaller than this required value. However, as
discussed in \S1, since a portion of the Reynolds layer may reside within
the H~I disk, it need not be sustained solely by the radiation escaping the
disk.

Throughout this paper, we have assumed that the formation rate of OB
associations is constant in time. For starburst galaxies, the observed star
formation rate is much higher than the time-averaged value. Therefore,
since $\etaesc(t)$ is at a maximum shortly after the formation of an
OB association, starburst galaxies having a morphology similar to our
Galaxy should have higher values of \fesc\ than those predicted for our
Galaxy. In addition, starburst galaxies may have lower H~I column
densities and lower gravitational fields, also contributing to a higher
escape fraction of ionizing radiation.  These considerations will be the
focus of our subsequent work.

If OB associations are indeed the sources of radiation responsible for
ionizing the Reynolds layer, it is possible that an appreciable amount of
radiation can escape the Galaxy entirely and contribute to the
intergalactic radiation field (\cite{bechtold87a,giroux97a}). If the
average escape fraction from low-redshift, star-forming galaxies exceeds
$\sim 5$\%, then the contribution of ionizing radiation to the
intergalactic medium by galaxies can exceed the contribution for QSOs
(\cite{shull95b,madau96a,giallongo97a}). We note, however, that our models
predict only the flux of radiation penetrating the Galactic halo, before a
portion of it is absorbed by the portion of the Reynolds layer situated
high above the Galactic plane.  In order to predict the fraction of
radiation that escapes the Galaxy entirely, we need a self-consistent model
of the density distribution of hydrogen gas within the halo, including the
contribution due to high velocity clouds, whose covering fraction is 
uncertain.  \cite*{murphy95} found that approximately 37\% of 21-cm sightlines 
near quasars showed HVCs with N$_{\rm HI} \geq 7\times10^{17}$ cm$^{-2}$. 
However, observations with the Parkes Multibeam Survey (\cite{putman99})
show that the spatial distribution of H~I in the HVCs is filamentary, 
suggesting that Lyman continuum could leak through.  Furthermore, the
actual HVC coverage fraction will vary considerably around the Galaxy,
depending on proximity to spiral arms and upwelling superbubbles. 

For specificity, if we assume $\fesc = 0.06$, then the upward ionizing flux 
is $\Phi_{\rm LyC} = 1.5 \times 10^6$ cm$^{-2}$ s$^{-1}$ above the H~I disk 
($Z\gta 1$ kpc).  If this flux is incident on the HVCs, and if we use 
the relationship between H$\alpha$ emission and ionizing flux, for 
optically thick clouds and photoionization equilibrium (\cite{bland-hawthorn99a}), 
then the predicted H$\alpha$ emission intensity is
\begin{equation}
I_E = 4.6\ {\rm mR} \left(\frac{\Phi_{\rm LyC}}{10^4\ {\rm cm}^{-2}\ {\rm
s^{-1}}}\right) = 0.69\ {\rm R}.
\end{equation}
This intensity is roughly a factor of two larger than the measured
intensities from the Magellanic Stream (\cite{weiner96a}) and
the Smith HVC (\cite{bland-hawthorn98a}). 

In summary, we have shown that the escape fraction of ionizing radiation
from star-forming regions of disk galaxies is strongly affected by the
presence of superbubbles and a cloudy ISM.  The shells of the expanding
superbubbles quickly trap or attenuate the ionizing flux, so that most of
the escaping radiation escapes shortly after the formation of the
superbubble. In our models, the escape fraction above 1 kpc range from
$\fesc \approx 3-20$\%. For a Gaussian SFH, \fesc\ is roughly a factor of
two lower than the results of \cite*{dove94b}, where superbubbles were not
considered.  If $\fesc \gta 8$\%, then the upward flux is sufficient to
sustain the Reynolds layer.  We have also examined the radiative
transfer in a two-phase (cloud/intercloud) medium.  We neglect the
filling factor of hot, low-density gas, consistent with recent O~VI
studies.  If a significant fraction of H~I is
distributed in cold clouds with $\nH \sim 30$ cm$^{-3}$ and column
densities $N_{\rm H} \sim (10^{19}-10^{20})$ cm$^{-2}$, $\fesc$ can be
reduced by a factor of $\sim 2-5$ if the clouds have a disk geometry. Thus,
the amount of ionizing photons that escape the disk into the IGM is likely
to be highly variable, and will depend on the luminosity distribution of OB
associations, the cloud structure, and the massive star formation history.
If the escaping LyC radiation is ever observed directly, it will confirm
these theoretical expectations but also verify the likelihood that stellar
ionizing radiation plays an important role in reionization of the
high-redshift IGM.

\medskip
\medskip
We thank Ralph Sutherland for the OB association luminosity curves.  We
thank Mark Giroux, Phil Maloney, John Bally, and Brad Gibson for informative
discussions.  We also thank the referee for useful suggestions that led to
major improvements to the paper. This work was supported by the Astrophysical
Theory Program at the University of Colorado, through grants NAG5-4063 from NASA
and AST96-17073 from NSF.


\newpage

\begin{thebibliography}{}

\bibitem[\protect\astroncite{Banfi et~al.}{1993}]{banfi93a}
Banfi, M., Rampazzo, R., Chincarini, G., \& Henry, R. B.~C. 1993, A\&A, 280,
  373

\bibitem[\protect\astroncite{Basu, Johnstone, \& Martin}{1999}]{basu99a}
Basu, S., Johnstone, D., \& Martin, P. G. 1999, ApJ, 516, 843 
  
\bibitem[\protect\astroncite{Bechtold et~al.}{1987}]{bechtold87a}
Bechtold, J., Weymann, R.~J., Lin, Z., \& Malkan, M.~A. 1987, ApJ, 315, 180

\bibitem[\protect\astroncite{Bertschinger}{1986}]{bertschinger86a}
Bertschinger, E. 1986, ApJ, 304, 154

\bibitem[\protect\astroncite{Bland-Hawthorn et~al.}{1998}]{bland-hawthorn98a}
Bland-Hawthorn, J.~B.,  Veilleux, S., Cecil, G.~N., Putman, M.~E., Gibson,
B.~K., \& Maloney, P.~R. 1998, MNRAS, 299, 611

\bibitem[\protect\astroncite{Bland-Hawthorn \& Maloney}{1999}]{bland-hawthorn99a}
Bland-Hawthorn, J.~B., \& Maloney, P.~R.  1999, ApJ, 510, 33

\bibitem[\protect\astroncite{Blondin et~al.}{1998}]{blondin98a}
Blondin, J.~M., Wright, E.~B., Borkowski, K.~J.,  \& Reynolds, S.~P. 1998,
  ApJ, 500, 342

\bibitem[\protect\astroncite{Bregman \& Harrington}{1986}]{bregman86a}
Bregman, J.~N., \& Harrington, J.~P.  1986, ApJ, 309, 833

\bibitem[\protect\astroncite{Castor, McCray \& Weaver}{1975}]{castor75a}
Castor, J., McCray, R., \& Weaver, R.  1975, ApJ, 200, L107

\bibitem[\protect\astroncite{Couchman \& Rees}{1986}]{couchman86a}
Couchman, H. M.~P., \& Rees, M.~J.  1986, MNRAS, 221, 53

\bibitem[\protect\astroncite{Dickey \& Lockman}{1990}]{dickey90a}
Dickey, J.~M., \& Lockman, F.~J.  1990, ARA\&A, 28, 215

\bibitem[\protect\astroncite{Dove \& Shull}{1994a}]{dove94a}
Dove, J.~B., \& Shull, J.~M.  1994a, ApJ, 423, 196

\bibitem[\protect\astroncite{Dove \& Shull}{1994b}]{dove94b}
Dove, J.~B., \& Shull, J.~M.  1994b, ApJ, 430, 222, DS94

\bibitem[\protect\astroncite{Dove, Ferrara, \& Shull}{1999}]{dove99b}
Dove, J.~B., Ferrara, A., \& Shull, J.~M. 1999, in preparation

\bibitem[\protect\astroncite{Elmegreen \& Lada}{1977}]{elmegreen77a}
Elmegreen, B.~G., \& Lada, C.~J.  1977, ApJ, 214, 725

\bibitem[\protect\astroncite{Fitzpatrick \& Spitzer}{1997}]{fitzpatrick97a}
Fitzpatrick, E.~L., \& Spitzer, L.~Jr. 1997, ApJ, 475, 623

\bibitem[\protect\astroncite{Garmany}{1998}]{garmany98a}
Garmany, C.~D.  1998,
\newblock in Origins, ed. C. Woodward, J.~M. Shull, H.~A.~Thronson, 
  ASP Conference Series, Vol. 148, 184

\bibitem[\protect\astroncite{Giallongo, Fontana \& Madau}{1997}]{giallongo97a}
Giallongo, E., Fontana, A., \& Madau, P.  1997, MNRAS, 289, 629

\bibitem[\protect\astroncite{Giroux \& Shull}{1997}]{giroux97a}
Giroux, M.~L., \& Shull, J.~M.  1997, AJ, 113, 1505

\bibitem[\protect\astroncite{Heiles}{1979}]{heiles79a}
Heiles, C. 1979, ApJ, 229, 533

\bibitem[\protect\astroncite{Kennicutt, Edgar \& Hodge}{1989}]{kennicutt89a}
Kennicutt, R.~C., J., Edgar, B.~K., \& Hodge, P.~W.  1989, ApJ, 337, 761

\bibitem[\protect\astroncite{Kompaneets}{1960}]{kompaneets60a}
Kompaneets, A.~S.  1960, Sov. Phys. Dokl., 5, 46

\bibitem[\protect\astroncite{Koo \& McKee}{1992}]{koo92a}
Koo, B.-C., \& McKee, C.~F.  1992, ApJ, 388, 93

\bibitem[\protect\astroncite{Lockman, Hobbs, \& Shull}{1986}]{lockman86}
Lockman, F.~J., Hobbs, L.~M., \& Shull, J.~M. 1986, ApJ, 301, 380 

\bibitem[\protect\astroncite{Mac~Low \& Ferrara}{1999}]{maclow99a}
Mac~Low, M.-M.,  \& Ferrara, A.  1999, ApJ, 513, 142

\bibitem[\protect\astroncite{Mac~Low \& McCray}{1988}]{maclow88a}
Mac~Low, M.-M.,  \& McCray, R.  1988, ApJ, 324, 776

\bibitem[\protect\astroncite{Mac~Low, McCray \& Norman}{1989}]{maclow89a}
Mac~Low, M.-M., McCray, R., \& Norman, M.~L. 1989, ApJ, 337, 141

\bibitem[\protect\astroncite{MacLow \& Norman}{1993}]{maclow93a}
MacLow, M.-M.,  \& Norman, M.~L.  1993, ApJ, 407, 207

\bibitem[\protect\astroncite{Madau \& Shull}{1996}]{madau96a}
Madau, P.,  \& Shull, J.~M.  1996, ApJ, 457, 551

\bibitem[\protect\astroncite{Massey}{1998}]{massey98a}
Massey, P., 1998, in New Views of the Magellanic Clouds, IAU Symp. 190,
eds. Chu, Y.-W. et al., p. 14

\bibitem[\protect\astroncite{Mathis}{1986}]{mathis86a}
Mathis, J.~S.  1986, ApJ, 301, 423

\bibitem[\protect\astroncite{McCray \& Kafatos}{1987}]{mccray87a}
McCray, R., \& Kafatos, M.  1987, ApJ, 317, 190

\bibitem[\protect\astroncite{McCray \& Snow}{1979}]{mccray79a}
McCray, R., \& Snow, T. P.  1979, ARA\&A, 17, 213

\bibitem[\protect\astroncite{McKee}{1986}]{mckee86a}
McKee, C.~F.  1986, Ap\&SS, 118, 383

\bibitem[\protect\astroncite{McKee}{1997}]{mckee97a}
McKee, C.~F., \&  Williams, J.~P.  1997, ApJ, 476, 144

\bibitem[\protect\astroncite{Mezger}{1978}]{mezger78a}
Mezger, P.~G.  1978, A\&A, 70, 565

\bibitem[\protect\astroncite{Miller \& Cox}{1993}]{miller93a}
Miller, W.~W., I., \& Cox, D.~P.  1993, ApJ, 417, 579

\bibitem[\protect\astroncite{Mineshige, Shibata \&
Shapiro}{1993}]{mineshige93a}
Mineshige, S., Shibata, K.,  \& Shapiro, P.~R. 1993, ApJ, 409, 663

\bibitem[\protect\astroncite{Murphy, Lockman, \& Savage}{1995}]{murphy95}
Murphy, E.~M., Lockman, F.~J., \& Savage, B.~D. 1995, ApJ, 447, 642

\bibitem[\protect\astroncite{Newman et~al.}{1999}]{newman99a}
Newman, W.~I., Symbalisty, E. M.~D., Ahrens, T.~J., \& Jones, E.~M.  1999,
Icarus, 138, 224

\bibitem[\protect\astroncite{Ostriker \& Cowie}{1981}]{ostriker81a}
Ostriker, J.~P.,  \& Cowie, L.~L. 1981, ApJ, 243, L127

\bibitem[\protect\astroncite{Ostriker \& McKee}{1988}]{ostriker88a}
Ostriker, J.~P.,  \& McKee, C.~F.  1988, Rev. Mod. Phys., 60, 1

\bibitem[\protect\astroncite{Putman \& Gibson}{1999}]{putman99}
Putman, M.~E., \& Gibson, B.~K. 1999, in Stromlo Workshop on High-Velocity
Clouds, ed. B.~K. Gibson \& M.~E. Putman, ASP Conf. Series, Vol 166, 276

\bibitem[\protect\astroncite{Reynolds}{1984}]{reynolds84a}
Reynolds, R.~J. 1984, ApJ, 282, 191

\bibitem[\protect\astroncite{Reynolds}{1991a}]{reynolds91a}
Reynolds, R.~J. 1991a, ApJ, 372, L17

\bibitem[\protect\astroncite{Reynolds}{1991b}]{reynolds91b}
Reynolds, R.~J.  1991b,
\newblock in The Interstellar Disk-Halo Connection in Galaxies, ed. H. Bloemen,
   (Dordrecht: Kluwer), ~67

\bibitem[\protect\astroncite{Rozas, Beckman \& Knapen}{1996}]{rozas96a}
Rozas, M., Beckman, J.~E.,  \& Knapen, J.~H.  1996, A\&A, 307, 735

\bibitem[\protect\astroncite{Shull \& Saken}{1995}]{shull95a}
Shull, J.~M.,  \& Saken, J.~M.  1995, ApJ, 444, 663

\bibitem[\protect\astroncite{Shull}{1995}]{shull95b} Shull, J, M.  1995,
\newblock in Airborne Astronomy Symposium on the Galactic Ecosystem: From Gas
  to Stars to Dust, ed. M.~R. Haas, J.~A. Davidson, E.~F. Erickson,
  ASP Conf. Series, Vol.~73, 365

\bibitem[\protect\astroncite{Stewart}{1998}]{stewart98a}
Stewart, S.~G.  1998, IAU Circ.~5302

\bibitem[\protect\astroncite{Sutherland \& Shull}{1999}]{sutherland99a}
Sutherland, R.~S., \& Shull, J.~M.  1999,
\newblock in preparation

\bibitem[\protect\astroncite{Tomisaka}{1998}]{tomisaka98a}
Tomisaka, K.  1998, MNRAS, 298, 797

\bibitem[\protect\astroncite{Tomisaka \& Ikeuchi}{1986}]{tomisaka86a}
Tomisaka, K.,  \& Ikeuchi, S.  1986, Pub. Astr. Soc. Japan, 38, 697

\bibitem[\protect\astroncite{Vishniac}{1983}]{vishniac83a}
Vishniac, E.~J.  1983, ApJ, 274, 152

\bibitem[\protect\astroncite{Vishniac}{1994}]{vishniac94a}
Vishniac, E.~J.  1994, ApJ, 428, 186

\bibitem[\protect\astroncite{Voit}{1988}]{voit88a}
Voit, G.~M.  1988, ApJ, 331, 343

\bibitem[\protect\astroncite{Weaver et~al.}{1977}]{weaver77a}
Weaver, R., McCray, R., Castor, J., Shapiro, P., \& Moore, R.  1977, ApJ,
  218, 317

\bibitem[\protect\astroncite{Weiner \& Williams}{1996}]{weiner96a}
Weiner, B.~J. \& Williams, T.~B. 1996, AJ, 111, 1156
\end{thebibliography}
\end{document}